\documentclass{article}
\usepackage{authblk}

\title{Physics-aware global Rietveld refinement for high-energy X-ray diffraction microscopy with application to reconstructing intragranular orientation and strain fields}
\author[1]{C.\,K.\ Cocke}
\author[1,2]{E.\ Camacho}
\author[1]{S.\,F.\ Gorske}
\author[1]{K.\,T.\ Faber}
\author[1]{K.\ Bhattacharya\thanks{Corresponding author: Kaushik Bhattacharya, bhatta@caltech.edu}}
\affil[1]{Division of Engineering and Applied Science, California Institute of Technology, Pasadena, CA 91125, USA}
\affil[2]{Department of Mechanical Engineering, Massachusetts Institute of Technology, Cambridge, MA 02139, USA }

\usepackage[T1]{fontenc} 
\usepackage[utf8]{inputenc} 
\usepackage{graphicx} 
\usepackage[top=1in,left=1in,right=1in,bottom=1in]{geometry} 
\usepackage{microtype} 
\usepackage[dvipsnames]{xcolor} 
\usepackage[toc,page]{appendix} 
\usepackage{float}
\usepackage{bm}

\usepackage{amsmath,amsfonts,amssymb,amsthm} 
\usepackage{mathtools} 
\usepackage{mathrsfs} 

\newcommand*\dif{\mathop{}\!\mathrm{d}} 
\DeclareMathOperator{\arctantwo}{arctan2}
\newcommand\hkl{{hk\ell}}
\DeclareMathOperator*{\argmin}{arg\,min}
\newcommand{\R}{\mathbb{R}}

\usepackage{tabularx} 
\usepackage{booktabs} 

\usepackage{caption}
\usepackage{hyperref}
\hypersetup{
    colorlinks=true,
    allcolors=blue,
    citecolor=ForestGreen,
    urlcolor=blue
}
		
\usepackage[capitalise,sort&compress]{cleveref}
\crefname{equation}{Equation}{Equations}
\crefname{figure}{Figure}{Figures}
\crefname{table}{Table}{Tables}
\crefname{algocf}{Algorithm}{Algorithms}

\Crefname{equation}{Equation}{Equations}
\Crefname{figure}{Figure}{Figures}
\Crefname{table}{Table}{Tables}
\Crefname{algocf}{Algorithm}{Algorithms}

\usepackage[style=ieee, citestyle=numeric-comp, maxbibnames=99, mincitenames=1, maxcitenames=2, uniquelist=false, uniquename=init, giveninits=true, isbn=false, url=false, eprint=false, clearlang=true, date=year, dashed=false]{biblatex} 
\DefineBibliographyStrings{english}{andothers = {et\addabbrvspace al\adddot}} 
\renewrobustcmd*{\bibinitdelim}{\,} 
\AtEveryBibitem{\clearlist{language}}
\AtEveryBibitem{\clearfield{note}}
\AtEveryBibitem{\clearfield{number}} 
\renewbibmacro{in:}{} 

\DeclareFieldFormat[article,periodical]{volume}{#1}
\DeclareFieldFormat[article,periodical]{pages}{#1}

\begin{document}

\maketitle

\begin{abstract}
    High-energy X-ray diffraction microscopy (HEDM) has emerged as a critical technique for studying the microstructure and, increasingly, strain fields in solids. However, current algorithmic or experimental methods to obtain intragranular fields are time-intensive, provide limited spatial resolution, or yield stress and strain fields that do not satisfy the universal laws of deformation (compatibility and equilibrium). In the context of standard HEDM, a novel physics-aware approach is presented in which the physics of deformation is included in the forward diffraction simulation to ensure that the reconstructed fields are physically meaningful. The entire simulated and experimental diffractograms are compared with a differentiable optimal transport--type objective, and a Rietveld refinement is carried out globally on the internal fields and grain topology using gradient-based optimization. The method is developed, verified with synthetic data, and demonstrated experimentally using near-field HEDM data from aluminum oxynitride (a brittle ceramic), with a reference implementation released as {\texttt{PARA-X}}. The reconstructions show remarkable improvement over existing methods (improved completeness and loss), and the high-fidelity, high-resolution recovery paves the way for using HEDM to study fine-scale deformation mechanics over large polycrystalline volumes.
\end{abstract}

\textbf{Keywords}: three-dimensional X-ray diffraction (3DXRD); high-energy X-ray diffraction microscopy (HEDM); intragranular strain; gradient-based optimization; optimal transport distance; grain boundary refinement; physics-aware optimization; polycrystalline Rietveld refinement

\newpage

\section{Introduction}

Since the 1990s, high-energy X-ray diffraction microscopy (HEDM), also known as three-dimensional X-ray diffraction (3DXRD), has become a common tool for nondestructive characterization of polycrystalline materials. Alongside infrastructure and experimental standardization of synchrotron and laboratory X-ray sources, there has been a concomitant maturation of the software and algorithms used to recover the state of the interrogated sample from diffraction measurements, a process normally termed ``reconstruction.'' Examples of established software include HEXRD \cite{bernierHEXRDHexrdRelease2023}, MIDAS \cite{sharmaFastMethodologyDetermine2012,sharmaFastMethodologyDetermine2012a}, and ImageD11 \cite{zotero-item-2884}, each with its own implementation, but based on similar ideas. These approaches follow the rotating crystal method, wherein the sample is illuminated by a collimated, monochromatic X-ray beam and rotated. As lattice planes of crystallites inside the sample locally satisfy the Bragg condition, associated coherent diffraction events are measured by downstream near- and far-field detectors. In the standard far-field processing approach, there are two distinct steps: indexing and fitting. The indexing step is a data-labeling process in which spots are fit to find their centroids and assigned an $\hkl$ reflection and parent grain label. For each grain, a subproblem is solved to optimize the grain-averaged orientation, strain, and position to minimize the distance between the indexed spots and those computed from forward diffraction \cite{lauridsenTrackingMethodStructural2001,bernierFarfieldHighenergyDiffraction2011}. In the standard near-field approach, spatial points are processed one-by-one with a Monte Carlo orientation search to maximize the overlap of all simulated reflections with the data \cite{suterForwardModelingMethod2006,liAdaptiveReconstructionMethod2013,nygrenAlgorithmResolvingIntragranular2020}.

The combination of near- and far-field modalities, coupled with their respective canonical reconstruction approaches, has led HEDM to become a standard tool for nondestructive 3D characterization. This has enabled studies on a wide range of materials, phenomena, and time scales, including plastic slip \cite{paganConnectingHeterogeneousSingle2014}, fatigue evolution \cite{limGrainReorientationStressstate2022}, phase transformations \cite{bucsekMeasuringStressinducedMartensite2018}, earth materials \cite{hurleyCrystallographicTextureStructure2025}, ferroelastic twinning \cite{bucsekFerroelasticTwinReorientation2019}, creep \cite{beaudoinBrightXraysReveal2017}, static and fatigue fracture \cite{gorskeInsituVisualizationGrowing2026a,spearThreedimensionalCharacterizationMicrostructurally2014}, among many others \cite{bernierHighEnergyXRayDiffraction2020}. The main limitation of these canonical analysis methodologies is that the strain field cannot be resolved intragranularly, limiting insights into the most revealing microstructural behaviors, which usually occur locally near grain boundaries.

The limited strain information provided by standard HEDM methods has led to the recent development of new methods and algorithms to attempt to resolve these intragranular features. On the experimental side, pencil-beam-based ``scanning'' or ``point-focused'' HEDM/3DXRD \cite{hayashiIntragranularThreedimensionalStress2019,hayashiScanningThreeDimensionalXray2023,henningssonMicrostructureStressMapping2024,zhangUnveiling3DSubgrain2025} has been developed to address this, along with corresponding reconstruction methodologies \cite{henningssonIntragranularStrainEstimation2021,henningssonReconstructingIntragranularStrain2020,liResolvingIntragranularStress2023,hayashiScanningThreeDimensionalXRay2017}. These methods use a small point-focused beam that is horizontally and vertically rastered across the sample. For each scan, very small regions of each grain are illuminated, which enables similar ideas from standard far-field reconstruction to be used for computing intragranular strains and orientations at the resolution of the beam dimensions. The major limitation of pencil-beam approaches comes from the same fact that provides their resolution: only a 1D line in the sample is illuminated, so the additional local information comes at the expense of either drastically longer experimental times or smaller interrogation domains. Other methods like dark-field X-ray microscopy \cite{simonsDarkfieldXrayMicroscopy2015} or differential aperture X-ray microscopy \cite{yangDifferentialapertureXrayStructural2004,larsonThreedimensionalXrayStructural2002} obtain intragranular fields, but through fundamentally different experimental methods than the monochromatic rotating crystal method of standard HEDM.

An alternative pathway toward improving resolution without lengthy experiments is through the development of post-processing methods that impose mechanical constraints to attempt to recover the intragranular strain fields. Previous work has developed methods to recover full-field information by projection to divergence-free fields \cite{zhouImposingEquilibriumExperimental2022}, to divergence-free fields while considering plastic incompatibility \cite{naraganiInterpretationIntragranularStrain2021}, and to the fully compatible and divergence-free regime \cite{cockeRecoveringIntragranularStrain2025}. Unfortunately, the problems these methods solve are highly ill-posed, even in the ideal fully elastic case \cite{cockeRecoveringIntragranularStrain2025}, so they cannot recover information lost by standard reconstruction methods. Further, they inherit any uncertainty in the original reconstruction, so the errors of each method compound.

In practice, the desire to probe at higher resolutions is at odds with experiment duration, and the significant costs of synchrotron facility operations limit the amount of data that can be obtained temporally, spatially, or both. The major objective from a method-development standpoint is to illuminate as large of a volume as possible while taking measurements in the shortest time possible. In this sense, one seeks to develop methods that still use the larger box or line-focused beams from standard near- and far-field HEDM, but provide additional intragranular information. A key limitation of standard reconstruction methods is that intensity measurements go almost entirely unused, where near-field diffractograms are binarized and far-field diffractograms only use intensity for improved centroid position calibration. Prior work has used full intensity measurements, such as \citeauthor{shenVoxelbasedStrainTensors2020}~\cite{shenVoxelbasedStrainTensors2020}, who developed a multistage per-grain pointwise orientation and strain update scheme to fit near-field spot intensity distributions through a Kullback--Leibler metric. Similarly, \citeauthor{reischigThreedimensionalReconstructionIntragranular2020}~\cite{reischigThreedimensionalReconstructionIntragranular2020} developed the iterative tensor field (ITF) method, which reconstructs the internal orientation and strain fields by means of an iterative two-stage method that alternates between solving for grain shape and deformation. Although powerful, this method requires mandatory indexing and numerous ad hoc schemes for regularization, smoothing, and grain stitching. Both methods also rely on decoupled multi-stage schemes to improve robustness at computational expense, and grains are considered independently, so compatibility (both in deformation and diffraction signals) between neighboring grains is not enforced.

This work solves the HEDM reconstruction inverse problem directly: we search for the complete 3D micromechanical state of the polycrystal whose entire simulated diffractogram best matches the measured diffractogram. This is a shift in the philosophy of reconstruction methods in that we solve the problem \emph{globally} and do not consider grain subproblems. Succinctly, in a gradient-based manner, we i) compute a full forward diffraction simulation from the entire illuminated volume of the polycrystal; ii) compare the entire simulated and experimental diffractograms to each other, explicitly considering the entire spatial intensity distribution; iii) update all fields simultaneously while imposing known physics on the fields within the polycrystal. This is analogous to the original ideas by \citeauthor{rietveldProfileRefinementMethod1969}~\cite{rietveldProfileRefinementMethod1969}, but in the case of the full polycrystalline state. Comparisons of full simulated diffractograms are enabled by advances in forward diffraction simulators, or virtual diffractometers. Examples include those by \citeauthor{wongFrameworkGeneratingSynthetic2013}~\cite{wongFrameworkGeneratingSynthetic2013},  \citeauthor{dawsonVirtualDiffractometerCreating2023}~\cite{dawsonVirtualDiffractometerCreating2023}, and \citeauthor{henningssonXrd_simulator3DXray2023}~\cite{henningssonXrd_simulator3DXray2023}, where each method enables high-accuracy calculations of the diffractogram associated with a single point, grain, or total deformed polycrystal. While these models are not implemented to be differentiable, it is possible to compute analytic gradients using symbolic tools, as the expressions are compositions of closed-form functions. In this way, one can compute the sensitivity of a diffractogram with respect to the fields inside the polycrystal, which enables gradient-based optimization of the internal fields.

There are numerous departures from existing methods to enable this fundamentally different method of reconstruction, though we follow and extend on the core ideas from post-processing methods, full spot spatial--intensity comparisons, and forward diffraction simulations. First, to enable a gradient-based approach, the forward diffraction model and the objective function must be differentiable. This is achieved by analytically differentiating the forward model symbolically and selecting an objective function that is differentiable. In standard far-field HEDM reconstruction algorithms, the distance is chosen to be the L2-distance between all pairs of simulated and experimental spot centroids, and the problem is solved independently over each grain \cite{bernierHighEnergyXRayDiffraction2020}. This requires a priori knowledge of the point and lattice plane to which each experimental spot corresponds, which is obtained via an indexing procedure to match the simulated and experimental spots. This indexing step is a source of potential error and should ideally be avoided altogether. In the standard near-field Monte Carlo approach, the distance is chosen to be the completeness, which is a summation of a binary function over spots and, consequently, is not differentiable. Further, one minus the completeness is not a metric: it is neither symmetric nor positive definite, and it does not satisfy the triangle inequality. This is not ideal from an optimization point of view. Finally, neither distance nominally considers intensity, and diffraction spots from each grain or position are treated as independent of each other. Considering these limitations, we wish to leverage a differentiable distance that \emph{globally} compares diffractograms. The most natural choice is the L2-distance over the detector voxels, but this results in the pixel size introducing a discrete nature to the comparison, and further, non-overlapping spots have zero gradient, so the initial guess is required to be nearly perfect. A natural alternative choice of distances is the class of integral probability metrics that remain valid and informative when the supports of the two diffractograms to compare are disjoint. Here we use the so-called ``Sinkhorn'' loss \cite{cuturiSinkhornDistancesLightspeed2013}, which is a regularization of the optimal transport distance that may be efficiently computed. To our knowledge, \citeauthor{kacprzakLaueIndexingOptimal2024}~\cite{kacprzakLaueIndexingOptimal2024} were the first to apply an optimal transport--type distance to inverse problems in diffraction, where they used the Sinkhorn loss for a Laue neutron diffraction indexing scheme.

The use of gradient-based global methods also allows us to naturally embed the physics of deformation in the reconstruction.  Specifically, we restrict the solution space, through partial differential equation (PDE)-constrained optimization, to satisfy mechanical admissibility by enforcing that the strain field solves the equations of elasticity. This follows similar ideas of the post-processing method by \citeauthor{cockeRecoveringIntragranularStrain2025}~\cite{cockeRecoveringIntragranularStrain2025}, extended to operate directly on raw diffraction data (see also adjacent work by \citeauthor{henningssonIntragranularStrainEstimation2021}~\cite{henningssonIntragranularStrainEstimation2021} and \citeauthor{wihardjaConstitutiveRelationsImages2025}~\cite{wihardjaConstitutiveRelationsImages2025}). One may also envision a less-restrictive divergence-free constraint imposed and implemented using a projection method akin to that by \citeauthor{zhouImposingEquilibriumExperimental2022}~\cite{zhouImposingEquilibriumExperimental2022}, or a fully unconstrained problem that more closely follows the methods by \citeauthor{shenVoxelbasedStrainTensors2020}~\cite{shenVoxelbasedStrainTensors2020} and \citeauthor{reischigThreedimensionalReconstructionIntragranular2020}~\cite{reischigThreedimensionalReconstructionIntragranular2020}, though implemented globally here. All of these are possible within this framework.

Imposing known physics on the deformation requires highly accurate grain boundary topology, and pointwise optimization alone cannot move grain boundaries from their initial guess. The problem is non-convex, so reorienting an element to that of a neighboring grain results in a different local minimum that cannot be reached by standard gradient-based methods. It is with this motivation that we introduce a method to simultaneously evolve the grain boundaries using the same forward modeling approach. Here, we follow ``shape optimization'' methods from the topology optimization literature \cite{sokolowskiIntroductionShapeOptimization1992,allaireStructuralOptimizationUsing2004}. The main idea of these methods is that one may obtain the sensitivity of the diffraction objective function with respect to the grain boundary positions rather than the pointwise fields. Given computed sensitivities, we then follow standard methods in grain growth and incorporate an advection method based on multiple level sets that evolves boundaries using the diffraction objective rather than through a physical process.

In what follows, we provide the core details of our gradient-based reconstruction algorithm in \cref{sec:methods}. Quantitative evaluation of the method using synthetic data with a known ground truth is given in \cref{sec:synthetic_examples}. We then apply the method to real experimental data collected from a compressed aluminum oxynitride ceramic polycrystal in \cref{sec:experimental_example}. Finally, considerations for future experiments and directions of this approach are discussed in \cref{sec:discussion}. 

\section{Methods}\label{sec:methods}

\subsection{Forward model}

\subsubsection{Abstracted diffraction}\label{sec:abstracted_diffraction}

For the purposes of the methods developed here, we may abstract the representation and modeling of diffraction. This abstract form allows one to implement arbitrary physics into the forward diffraction model, so long as the resulting model satisfies the assumptions that follow. Nonetheless, we extensively describe the forward kinematic diffraction model used in this work in \cref{sec:forward_diffraction_model}, which follows standard expressions for rotating crystal diffraction. 
\begin{figure}
    \centering
    \includegraphics[width=\textwidth]{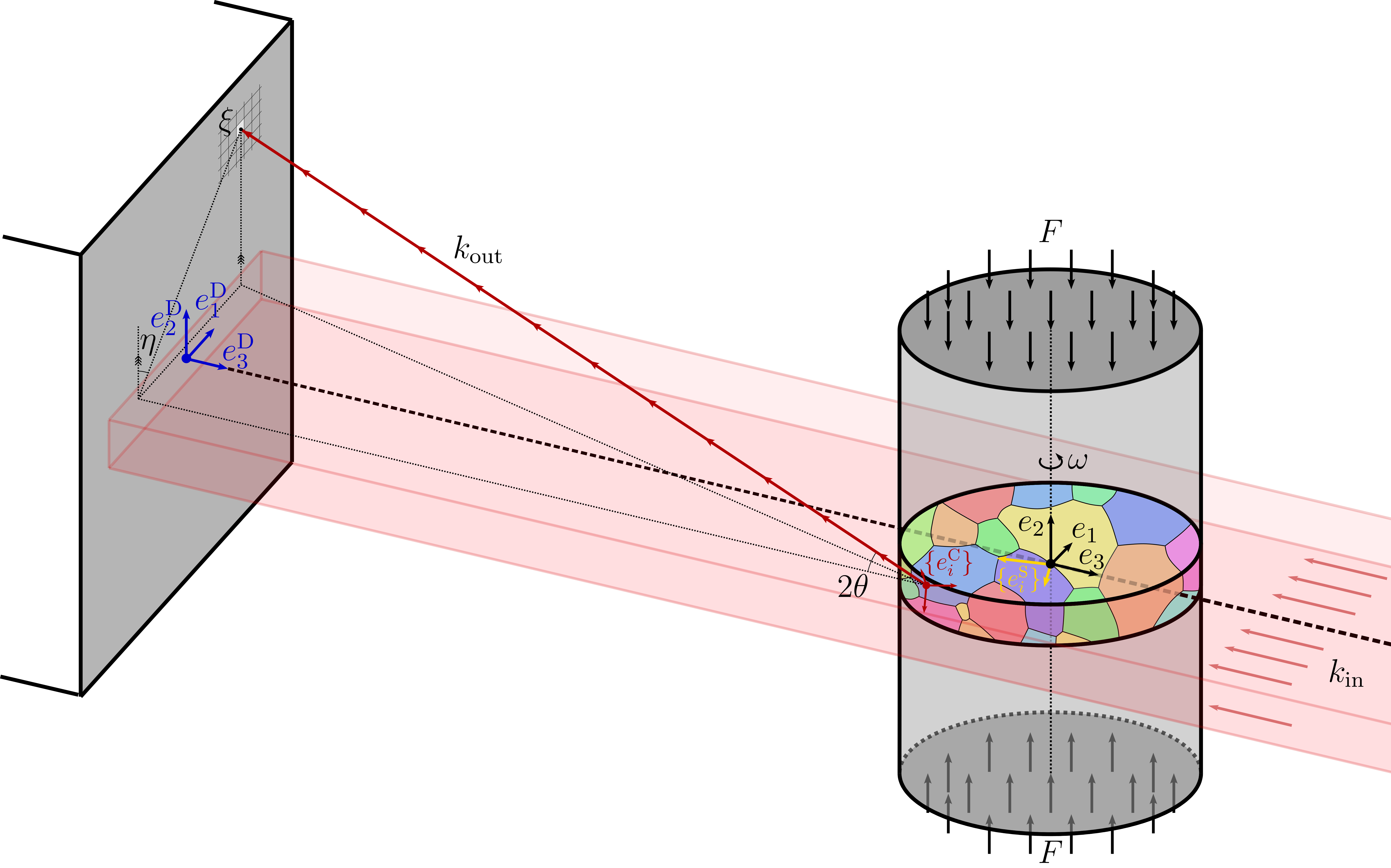}
    \caption{Diagram of a near-field HEDM scan. The coordinate system follows the HEXRD convention \cite{bernierHEXRDHexrdRelease2023}.}
    \label{fig:coordinate_system}
\end{figure}
We begin by following the standard rotating crystal method used during HEDM experiments, where, as the sample rotates continuously about a single axis through $\omega \in [-\pi, \pi]$, diffraction events are measured at normalized positions $\xi \in [-0.5, 0.5]^2$ on a 2D downstream detector, as shown in \cref{fig:coordinate_system}. We assume that diffraction is local, where a crystal oriented at $q \in \mathrm{SO}(3)$ with respect to the sample frame, subjected to the displacement gradient $h \in {\mathbb R}^{3\times 3}$, and located at position $x \in \mathbb R^3$, gives rise to a 3D diffractogram\footnote{Ideally, a diffractogram is a collection of discrete spots, i.e., a measure consisting of a sum of Dirac masses. However, in reality, a variety of experimental factors regularize it, so we assume here that it is absolutely continuous with respect to the Lebesgue measure.} (image) or intensity distribution 
\begin{equation}
I_0(\xi, \omega; x, q, h) = f(\xi, \omega, x, q, h).
\end{equation}
An exact definition of the function $f$ used in this work is given in \cref{eq:diffraction_forward_function}; for convenience, we may assume that $f$ is smooth (continuously differentiable) in $q$ and $h$. To simplify the notation in further expressions, we denote $\mathcal Y$ as the ``detector--rotation space,'' i.e., the product space of $\xi$ and $\omega$:
\begin{equation}\label{eq:detector_rotation_space}
    \mathcal Y \coloneqq \{(\xi, \omega) \mid \xi \in [-0.5, 0.5]^2, \ \omega \in [-\pi, \pi]\},
\end{equation}
and let $y \coloneqq (\xi, \omega) \in \mathcal Y$ be the detector--rotation position in this product space. 

In the case of a full polycrystal, we collect data from a sampling volume $\Omega \subset {\mathbb R}^3$. We assume that our sample has an orientation field described by a map $q \colon \Omega \to \mathrm{SO}(3)$ where $q(x)$ denotes the orientation of the crystal lattice at position $x$ in the specimen\footnote{Note $q(x)$ is piecewise constant for an ideal polycrystal.}. Further, residual or load-dependent deformation in the crystal leads to a displacement gradient field $h \colon \Omega \to \mathbb{R}^{3 \times 3}$. The entire diffractogram from the sampling volume $\Omega$ is then obtained by superposing the diffractograms of each point in the sample:
\begin{equation}\label{eq:diff}
I(y) = \int_\Omega f (y, x, q(x), h(x)) \dif x.
\end{equation}
Note that this is a functional of the fields $q$ and $h$. 

\subsubsection{Elasticity}\label{sec:elasticity}

\begin{figure}
    \centering
    \includegraphics[width=\textwidth]{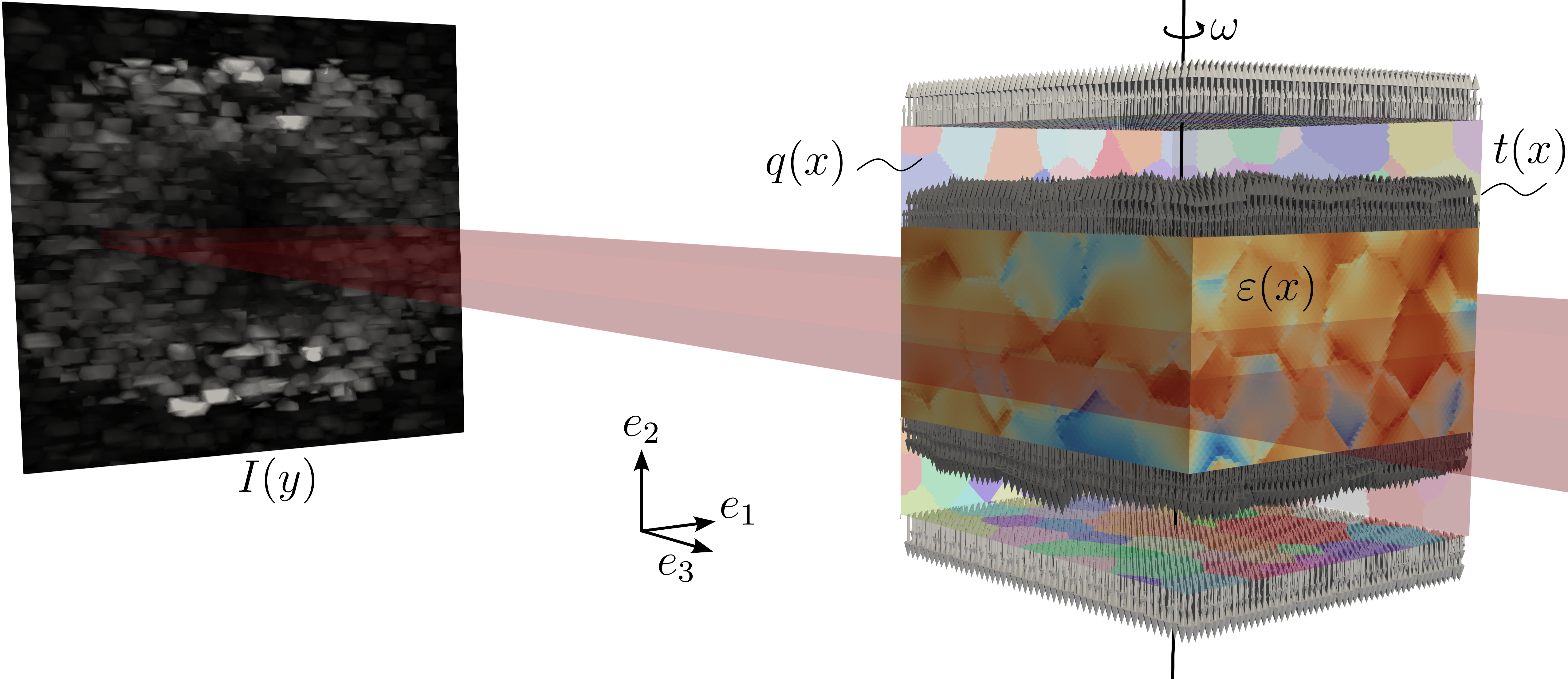}
    \caption{Diagram of orientation and strain fields producing a resultant near-field diffractogram. The axial strain field is shown in the interrogated volume with the rest of the microstructure semi-transparent. The applied loads at the sample boundary may be assumed uniform, but the internal loads on the measurement boundary may not be. The rotation-summed synthetic near-field diffractogram for the layer is shown; the far-field detector is not shown.}
    \label{fig:synthetic_data_generation}
\end{figure}

In this work, we are specifically interested in elastic deformations of brittle materials under an applied axial load with net force $F_\mathrm{net}$. Our sampling volume is typically cylindrical with a cross-section $C$ and height $H$, $\Omega = C \times (-H/2,H/2)$, and is subjected to a traction $t(x)$ on the top and bottom surfaces $\partial_\mathrm{tb} \Omega = \partial_\mathrm{t} \Omega \cup \partial_\mathrm{b} \Omega$ where $\partial_\mathrm{t} \Omega = C \times \{H/2\}, \partial_\mathrm{b} \Omega= C \times \{-H/2\}$, while it is traction-free on the lateral surface $\partial \Omega \setminus \partial_\mathrm{tb}\Omega$. The traction distribution must satisfy the conditions
\begin{equation} \label{eq:self_equilibrium}
    \int_{\partial_\mathrm{t}\Omega} t_2 \dif A = -\int_{\partial_\mathrm{b}\Omega} t_2 \dif A = F_\mathrm{net}, \int_{\partial_\mathrm{tb}\Omega} t \dif A = 0, \int_{\partial_\mathrm{tb}\Omega} x \times t \dif A = 0,
\end{equation}
so that it is self-equilibrated and consistent with the measured load applied on the specimen (following \cref{fig:coordinate_system}, loading is applied in the $e_2$ direction). Given $t \in L^2 (\partial_\mathrm{tb} \Omega; \mathbb R^3)$ satisfying \cref{eq:self_equilibrium}, we can obtain the displacement field by solving the elasticity problem in the sampling volume\footnote{We have assumed for simplicity that there is no residual stress or strain. This can easily be incorporated into our formulation by adding an unknown residual field and optimizing over this field as well.},
\begin{equation}\label{eq:elasticity_pde}
\begin{cases}
    \nabla \cdot \bigl(\mathbb C (q(x)) \nabla u(x)\bigr) = 0 \quad &\text{ in } \Omega, \\
    \bigl(\mathbb C(q(x)) \nabla u(x)\bigr) \cdot n(x)  = t(x) & \text{ on } \partial_\mathrm{tb} \Omega.
\end{cases}
\end{equation}
We can now obtain the displacement gradient field to be $h(x) = \nabla u(x)$. We substitute this into \cref{eq:diff}, and the diffractogram is 
\begin{equation}\label{eq:full_forward_diffractogram}
I_\mathrm{sim}(y) = \int_\Omega f (y, x, q(x), \nabla u(x)) \dif x,
\end{equation}
where we use the subscript ``sim'' to signify that this is the simulated diffractogram obtained using the forward diffraction model. In summary, $I_\mathrm{sim}$ is a functional of the crystallographic orientation field $q$ and the internal traction distribution $t$. Given known $q$ and $t$, we solve \cref{eq:elasticity_pde} for the displacement field and can compute the diffractogram $I_\mathrm{sim}$. Following the linearized elasticity problem in \cref{eq:elasticity_pde}, we henceforth assume strain and elastic strain to be synonymous, with $\varepsilon = \operatorname{sym} \nabla u$. A visualization of how the diffractogram and orientation, strain, and traction fields are related is shown in \cref{fig:synthetic_data_generation}.

\subsection{Inverse problem}\label{sec:inverse_problem}

Given a specimen with net applied force $F_\mathrm{net}$, we seek to ``reconstruct'' the crystallographic orientation and strain fields from the measured diffractogram over the sampling volume. We do so by comparing the simulated diffractogram $I_\mathrm{sim}$, computed by solving the forward problem, with the experimentally observed diffractogram $I_\mathrm{exp}$. We formulate this as an optimization problem, one of minimizing a loss function defined as some distance between the forward simulated and experimental diffractograms:
\begin{equation}\label{eq:general_inverse_problem}
(\bar{q}, \bar{t}) = \argmin_{q \in \mathcal{Q}, t \in \mathcal{T}} \ d(I_\mathrm{sim}, I_\mathrm{exp}),
\end{equation}
where 
\begin{equation}
\mathcal{Q} = L^2(\Omega; \mathrm{SO}(3)), \quad \mathcal{T} = \{t \in L^2(\partial_\mathrm{tb} \Omega; {\mathbb R}^3) \mid t \text{ satisfies \cref{eq:self_equilibrium}} \} .
\end{equation}
This is an indirect inverse problem, i.e., a PDE-constrained optimization problem. The forward simulated diffractogram $I_\mathrm{sim}$, and therefore the loss function, depends on the displacement field $u$ that is obtained by solving the elasticity PDE in \cref{eq:elasticity_pde} for a given $q \in \mathcal{Q}$ and $t \in \mathcal{T}$.

\subsubsection{Choice of distance}\label{sec:distance}

We choose an optimal transport distance to compare diffractograms. The standard Wasserstein metric is infeasible to compute for large point clouds; however, regularized approximations of the Wasserstein metric exist that are both differentiable and feasible to compute. Specifics regarding such regularizations can be found elsewhere \cite{feydyGeometricDataAnalysis2020,peyreComputationalOptimalTransport2020}. Here, we choose the distance function to be the unbalanced Sinkhorn divergence \cite{sejourneSinkhornDivergencesUnbalanced2023} between the simulated and experimental diffractograms:
\begin{equation}
    d \coloneqq S_{\epsilon,\rho},
\end{equation}
where $\sqrt{\epsilon}$ and $\sqrt{\rho}$ are the blur and reach, which are hyperparameters that control the amount of regularization between points and the region of influence, respectively. Any transport cost function may be used, though we use the standard L2-distance function. Note that $\nabla S_{\epsilon,\rho} \coloneqq d_{,I} \in C^1(\mathcal Y; \R)$ is well-defined, and that the Sinkhorn divergence $S_{\epsilon,\rho}$ is a valid divergence \cite{feydyGeometricDataAnalysis2020}.

\subsubsection{Multiple layers or detectors}

Multiple detector distances, rasterized scans, or varied beam topologies are naturally integrated within the method presented here by taking the objective to be the sum of all distance functions computed over all scans. In X-ray diffraction experiments, the beam size is typically much smaller than the region of interest one wishes to interrogate, so the beam is rastered (vertically, horizontally, or both) across the sample to illuminate the entire region of interest. This summed objective allows for any arbitrary scan sequences to be used simultaneously during the reconstruction, making this methodology applicable to all HEDM-based measurements (e.g., rastered line- or point-focused beams). 

\subsection{Pointwise optimization}\label{sec:pde_constrained_point_wise}

Given the PDE-constrained optimization problem in \cref{eq:general_inverse_problem}, one may compute sensitivities of the orientation and traction fields by means of an adjoint method \cite{hinzeOptimizationPDEConstraints2009}. We define the outer variation using a subscript notation: $d_{,q} \coloneqq \mathrm d / \mathrm{d} s \, d(q + s \delta q)|_{s=0}$ where $q,\delta q \in \mathcal Q$ (note that we represent $q$ as a unit-quaternion field). Following the adjoint procedure, the sensitivity of the objective with respect to the orientation field, which includes the implicit strain-field dependency, is
\begin{equation}\label{eq:orientation_sensitivity}
     \delta_q d(\delta q) = d_{,q}(\delta q) + \int_\Omega \varepsilon(\psi) \cdot \left[\left(\frac{\partial \mathbb C}{\partial q}(q) \cdot \delta q \right) \varepsilon(u) \right] \dif x,
\end{equation}
where $\psi$ is the adjoint displacement and ${\partial \mathbb C}/{\partial q}$ may be analytically computed (using indicial notation, with $\kappa$ from $1,\ldots,4$) as
\begin{equation}
    \mathbb C_{ijkl,\kappa}(q) = (Q_{im,\kappa}Q_{jn}Q_{ko}Q_{lp} + Q_{im}Q_{jn,\kappa}Q_{ko}Q_{lp} + Q_{im}Q_{jn}Q_{ko,\kappa}Q_{lp} + Q_{im}Q_{jn}Q_{ko}Q_{lp,\kappa})\mathbb C_{mnop},
\end{equation}
where $Q(q)$ is the standard rotation matrix defined by the unit-quaternion $q$, and the sensitivity with respect to the applied traction distribution is
\begin{equation}\label{eq:traction_sensitivity}
     \delta_t d(\delta t) = -\int_{\partial_\mathrm{tb}\Omega}  \psi \cdot \delta t \dif S.
\end{equation}
\Cref{eq:orientation_sensitivity,eq:traction_sensitivity} hold given the forward elasticity problem
\begin{equation}\label{eq:forward_elastic_problem}
     \int_\Omega \varepsilon(\delta \psi) \cdot \mathbb C(q) \varepsilon(u) \dif x = \int_{\partial_\mathrm{tb}\Omega} \delta \psi \cdot t \dif S
\end{equation}
is solved for $u$, and the adjoint problem is solved for $\psi$:
\begin{equation}\label{eq:adjoint_problem}
     \int_\Omega \varepsilon(\delta u) \cdot \mathbb C(q) \varepsilon(\psi) \dif x = -\int_\Omega\int_{\mathcal Y} d_{,I}(y) f_{,u}(y, x; \nabla \delta u) \dif y \dif x,
\end{equation}
where the integrals may be exchanged by the fact that $d_{,I} \in C^1(\mathcal Y; \mathbb R)$ and $f$ is assumed smooth.

The discrete gradients are given by solving \cref{eq:forward_elastic_problem,eq:adjoint_problem} with a standard finite element method, and \cref{eq:orientation_sensitivity,eq:traction_sensitivity} are subsequently computed discretely for each element. Given the computed gradients, the orientation and traction fields are updated with gradient-based optimization. Here we use projected gradient descent without line search for convenience, though this directly extends to other gradient-based optimization algorithms. For the orientations, we represent them as unit-quaternions, so the feasible set is the unit 3-sphere; hence, projection to the feasible set is simply quaternion normalization. For traction updates, we compute the discrete version of the constraint in \cref{eq:self_equilibrium} and project the update onto the feasible set (see \cref{sec:traction_discrete_constraint}). 

\subsection{Grain boundary optimization}\label{sec:grain_boundary_optimization}

Due to the nonlocal nature of the elliptic linear-elastic PDE solution operator, it is necessary to capture the grain structure as accurately as possible to reduce error in the entire strain field. The pointwise optimization introduced in \cref{sec:pde_constrained_point_wise} is unable to complete the large rotations necessary to update a single point from one grain orientation to another, so additional methods are required. Therefore, we introduce a method to evolve the grain boundaries using the same forward modeling approach, where we adopt ideas from numerical methods of grain growth, though the driving force for grain boundary evolution is determined by using the diffraction objective rather than by some physical energy.

\subsubsection{Inner variation}

We now compute the inner variation of the objective function, which quantifies how much the grain boundaries have to be moved to minimize the objective. We search over grain boundary perturbations $v \in \mathcal V$ with 
\begin{equation}\label{eq:velocity_perturbation_space}
    \mathcal V \coloneqq \{v \in H^1(\Omega; \R^3) \mid v \cdot n = 0 \ \text{on} \ \partial \Omega\},
\end{equation}
where the normal constraint in \cref{eq:velocity_perturbation_space} removes any perturbations in directions that cause material deformation outside of the domain. We usually solve the reconstruction on some known domain (e.g., from computed tomography), so this constraint is natural. Taking $v \in \mathcal V \cap W^{1,\infty}$, then the inner variation of the objective\footnote{For simplicity, this does not consider the total derivative or the PDE-constrained dependency of the strain field on the grain boundary positions; this can be considered through shape differentiation with an adjoint method \cite{allaireStructuralOptimizationUsing2004}.}, denoted by $\delta_\mathrm{in}d$, is
\begin{equation}\label{eq:objective_inner_variation}
    \delta_\mathrm{in}d(v) = \int_{\mathcal Y} d_{,I}(y)\delta_\mathrm{in}I(y, v) \dif y.
\end{equation}
The inner term, which is essentially the perturbation of the diffractogram with respect to grain boundary positions, may then be evaluated from \cref{eq:full_forward_diffractogram} as
\begin{equation}\label{eq:diffractogram_inner_variation}
   \delta_\mathrm{in} I(y, v) = \int_{\Omega} \left[\left.\frac{\partial f}{\partial x}\right|_{q,\nabla u} \cdot v + f \operatorname{div}v\right] \dif x,
\end{equation}
where ${\partial f}/{\partial x}$ is explicitly evaluated holding $q$ and $\nabla u$ fixed. We then substitute \cref{eq:diffractogram_inner_variation} into \cref{eq:objective_inner_variation} to get the final form of the inner variation:
\begin{equation}\label{eq:final_inner_variation}
     \delta_\mathrm{in}d(v) = \int_{\Omega} \int_{\mathcal Y} d_{,I}(y) \left[\frac{\partial f}{\partial x}(y,x) \cdot v(x) + f(y,x) \operatorname{div}v(x) \right] \dif y \dif x,
\end{equation}
where we have again used the smoothness assumption on $f$ to exchange the integrals. 

\subsubsection{Virtual velocity field}

Given the inner variation in \cref{eq:final_inner_variation}, one still must determine a descent direction. The natural choice, which depends on the inner product on $\mathcal V$, is given by choosing $v \in \mathcal V$ which solves $\langle v, \delta v\rangle_{\mathcal V} = -\delta_\mathrm{in} d(\delta v)$\footnote{With this choice, the descent direction decreases the objective: $\delta_\mathrm{in} d(v) = -\langle v, v\rangle_{\mathcal V} \le 0$.}, i.e.\
\begin{equation}\label{eq:helmholtz_velocity_field}
    \int_\Omega (\alpha^2\nabla v\cdot \nabla \delta v + v \cdot \delta v) \dif x = -\delta_\mathrm{in}d(\delta v).
\end{equation}
The parameter $\alpha$ is free and defines a length scale over which the velocity is smoothed. \Cref{eq:helmholtz_velocity_field} may be easily solved for $v(x)$ on the same mesh and elements as the elasticity problem, and the linear tangential constraint may be directly imposed. We call the obtained $v(x)$ the ``virtual velocity field,'' which we then use to advect the grain boundaries over some virtual time $T$. The velocity field is virtual in the sense that it derives from a non-physical process, but it may still be thought of as a velocity by the grain boundary evolution methods described below.

\subsubsection{Level set advection}\label{sec:level_set_advection}

Following the ideas of  \citeauthor{osherFrontsPropagatingCurvaturedependent1988}~\cite{osherFrontsPropagatingCurvaturedependent1988}, we choose to use a level set method to evolve the grain boundaries given a computed virtual velocity field $v(x)$. Consider a particle at point $x \in \Omega$ on the zero level set of $\phi$, i.e., at the interface. As the level set evolves over time, the particle should remain on the interface:
\begin{equation}\label{eq:hamilton_jacobi_precursor}
    \frac{\mathrm{d}}{\mathrm{d}t}\phi(x(t), t) = 0.
\end{equation}
By evaluating \cref{eq:hamilton_jacobi_precursor}, the Hamilton--Jacobi equation must hold in the domain: 
\begin{equation}\label{eq:hamilton_jacobi_level_set}
\begin{cases}
    \frac{\partial \phi}{\partial t}(x, t) + v(x) \cdot \nabla \phi(x, t) = 0 & \text{in} \ \Omega \times \R^+,\\
    \phi(x, 0) = \phi_0(x) & \text{in} \ \Omega,
\end{cases}
\end{equation}
where $v = \mathrm{d}{x}/\mathrm{d} t$ is the virtual velocity field, which we take to be the descent direction $v$ computed in \cref{eq:helmholtz_velocity_field}. \Cref{eq:hamilton_jacobi_level_set} is unstable if solved with Galerkin finite elements (the natural choice for the rest of our methods), so instead we solve \cref{eq:hamilton_jacobi_level_set} using a semi-Lagrangian backward characteristic scheme \cite{strainSemiLagrangianMethodsLevel1999,allaireShapeOptimizationLevel2014}. In this approach, the level set at a given point $x$ and time $t$ is computed by integrating the characteristic backward in time under the prescribed velocity field. Here, we simply choose to do so with explicit Euler time discretization, so the level set may be evolved from the $n$th to $(n+1)$th step as
\begin{equation}
    \phi_{n+1}(x) = \phi_{n}\bigl(x - \Delta t v(x)\bigr),
\end{equation}
where $\Delta t = t_{n+1} - t_n$ is the virtual time increment. The level set $\phi$ may be easily evaluated at the arbitrary point $x - \Delta t v(x)$ through nodal interpolation with the finite element shape functions.

\subsubsection{Multiple level set approach}

We follow the multiple level set method from \citeauthor{merrimanMotionMultipleJunctions1994}~\cite{merrimanMotionMultipleJunctions1994} and \citeauthor{zhangMultipleLevelSet2008}~\cite{zhangMultipleLevelSet2008} and describe the grain boundary--partitioned domain with $N$ distinct level sets $\phi_i$, $i = 1,\ldots,N$, where $N$ is the number of grains, each describing the signed distance to the boundary of the associated grain (taking $\phi_i > 0$ inside grain $i$). Following \cref{sec:level_set_advection}, we evolve each level set independently up to some final virtual time $T$. After evolution, a corrector step is simultaneously applied between the level sets of all grains to patch overlapping or void regions \cite{zhangMultipleLevelSet2008}:
\begin{equation}
    \phi^\mathrm{c}_i = \frac{1}{2}(\phi_i - \max_{j \ne i} \phi_j),
\end{equation}
where $\phi^\mathrm{c}_i$ are the corrected values. The level sets are then reinitialized to be signed distance functions. Exact geometric distances are evaluated at the vertices of all elements containing a zero level set (grain boundary), and these vertices are then used to seed a fast marching method that updates the level set values over the rest of the tetrahedral mesh \cite{lelievreComputingFirstarrivalSeismic2011}. The use of a fast marching method here is critical for performance when compared to a fully geometric approach. After reinitialization, the maximum level set value is computed for all elements in the domain. If the element grain ID differs from the ID of the maximum level set, it is marked to be updated. Marked elements are then updated sequentially by copying the orientation from neighboring elements with the new grain ID. This update scheme preserves local mosaic spread near the boundary, whereas simply using the grain-averaged orientation does not.

Grains must be explicitly defined here in order to assemble the initial level sets. We compute grains by taking all connected\footnote{We take ``connected'' to mean elements that share vertices, not faces.} elements whose neighbor-to-neighbor misorientation is below a defined tolerance ($2.5^\circ$ in this work). It is possible for grains to be annihilated, but not created, by this advection process, so missing grains cannot be restored without modification to this framework.

\subsection{Computational details}

We now outline how the previously described methods are implemented in a computationally efficient manner. Further non-essential implementation details are given in \cref{sec:further_computational_details}.

\subsubsection{Discretization}\label{sec:discretization_of_diffraction}

Within the implementation of this framework, we store and compare diffractograms as empirical measures, i.e., as a sum of weighted Dirac masses. This is enabled by means of the Wasserstein metric (and Sinkhorn divergence) naturally handling discrete measures. Stronger measure-theoretic tools are necessary for analysis of variations of the forward diffractogram, which becomes a measure-valued functional, but the ideas here remain the same. In this implementation, the forward and experimental diffractograms $I_\mathrm{sim}$ and $I_\mathrm{exp}$ are represented as the Dirac sums (by means of the forward diffraction function defined in \cref{eq:diffraction_forward_function})
\begin{equation}\label{eq:discrete_probability_measures}
    \nu_\mathrm{sim} = \beta\sum_{i=1}^n I^\mathrm{sim}_i \delta_{y_i}, \quad \nu_\mathrm{exp} = \frac{1}{\sum_{j=1}^m I^\mathrm{exp}_j}\sum_{j=1}^m I^\mathrm{exp}_j \delta_{y_j},
\end{equation}
where $y_i,y_j \in \mathcal Y$, $I^\mathrm{sim}_i,I^\mathrm{exp}_j > 0$, $\delta_y$ is the Dirac delta function centered at point $y$, and $\beta$ is a normalization constant that makes $\nu_\mathrm{sim}$ approximately a probability measure (see \cref{sec:normalization}). The simulated diffractogram then contains all $n$ Dirac masses from each distinct diffraction event occurring in the simulation. For the $m$ non-zero pixels in the experimental diffractogram, a Dirac mass is assigned at the center of the detector--rotation ``voxel'' (i.e., centered at the pixel and rotation increment center) with weight equal to the intensity of the voxel.

Discretely, the distance $d(\nu_\mathrm{sim},\nu_\mathrm{exp}) \coloneqq S_{\epsilon,\rho}(\nu_\mathrm{sim},\nu_\mathrm{exp})$ and its gradients $\partial d / \partial y_{i}$ and $\partial d / \partial I_i$ are computed using the multiscale Sinkhorn algorithm implemented in the GeomLoss library \cite{feydyGeometricDataAnalysis2020}. We use graphics processing units (GPUs) for Sinkhorn loss calculations and do so with 32-bit floating-point precision. Further, since we generally use a strict value for the reach (unbalanced transport) due to a good initial guess, the transport is effectively local, so we begin the $\epsilon$-annealing schedule at $10 \sqrt{\rho}$. These factors lead to the evaluation of the optimal transport loss being numerically feasible for our problem, even as the diffractogram point clouds exceed millions of points (e.g., see \cref{fig:time_breakdown}). 

\subsubsection{Binning}\label{sec:binning}

\begin{figure}
    \centering
    \includegraphics[width=.6\textwidth]{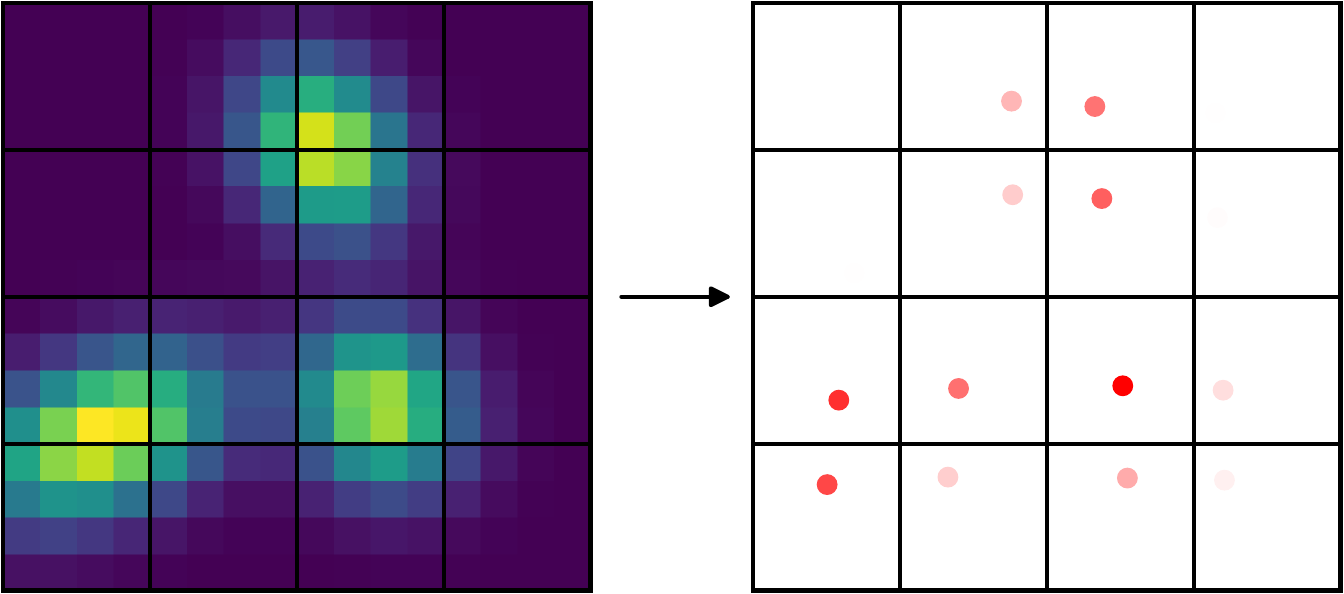}
    \caption{Spot binning example from a $16 \times 16$ 2D grid to a $4 \times 4$ 2D grid (the third rotation dimension is not shown), with the black lines denoting the bins. The positions of the red points are intensity-weighted averages of the pixel positions, so they do not lie in the centers of the bins.}
    \label{fig:spot_binning}
\end{figure}

In the forward diffraction model, the $n$-length spot and spot-gradient arrays are never explicitly constructed, which alleviates memory constraints when $n$ is large (e.g., for high-resolution microstructures or element sub-sampling) and decreases the computational cost of loss evaluation. Instead, all spots contained within a detector--rotation voxel are aggregated into a single point before evaluation of the Sinkhorn distance. This is the de facto standard method for virtual diffractometers, but the difference here is that we use the intensity-weighted center of mass within each bin rather than the actual center of the bin, e.g., see \cref{fig:spot_binning}. This allows for sub-pixel movement of diffraction spots and is also differentiable inside each bin without regularization. Temporarily adopting bold vector notation for clarity, the binning procedure follows standard clustering: at each voxel, a single spot is assigned with the total summed intensity $I^{\mathrm c}_i$, and the intensity-weighted center of mass, $\bm y^{\mathrm c}_{i}$, of all spots inside the voxel, $V_i$:
\begin{subequations}\label{eq:binning}
    \begin{equation}
        I^{\mathrm c}_i \coloneqq \sum_{k \in V_i} I_k,
    \end{equation}
    \begin{equation}\label{eq:binned_positions}
        \bm y^{\mathrm c}_{i} \coloneqq \frac{1}{I^\mathrm{c}_i} \sum_{k \in V_i} I_k \bm y_{k}.
    \end{equation}
\end{subequations}
Under this binning, the gradients of the binned spot positions are then evaluated as\footnote{One can also evaluate the gradient of the bin intensity; these gradients are neglected as discussed in \cref{sec:intensity}.}
\begin{equation}
    \frac{\partial \bm y^{\mathrm c}_{i}}{\partial (\cdot)} = \frac{1}{I^{\mathrm c}_i} \sum_{k \in V_i} I_k \frac{\partial \bm y_{k}}{\partial (\cdot)}.
\end{equation}
Given $n$ binned spots, the gradient of the distance function is then
\begin{align}
    \frac{\partial d}{\partial (\cdot)}  = \sum_{i=1}^{n} \frac{\partial d}{\partial \bm y^{\mathrm c}_{i}} \cdot \frac{\partial \bm y^{\mathrm c}_{i}}{\partial (\cdot)} 
     = \sum_{i=1}^{n}  \sum_{k \in V_i}\frac{I_k}{I^{\mathrm c}_i} \frac{\partial d}{\partial \bm y^{\mathrm c}_{i}} \cdot \frac{\partial \bm y_{k}}{\partial (\cdot)}. \label{eq:binned_gradient_update}
\end{align}

Binning is performed on-the-fly in a two-pass process. In the first pass, the sparse binned spots are computed and stored in a hash map by accumulating the intensity-weighted sum over each voxel following \cref{eq:binning}; during this process, the spot position gradients are not computed. After all diffraction events are computed, the binned spot positions are finalized by culling bins with total intensity less than some threshold, set by the dynamic range of the detector, and applying the normalizing prefactor in \cref{eq:binned_positions}. The objective function and its gradient with respect to the spots are then evaluated. In the second pass, the spots are recomputed, this time with their gradients, which are then used to update the local objective gradients via \cref{eq:binned_gradient_update}.

Binning may be performed at the standard pixel size and rotation increment; however, it is sometimes necessary for computational tractability of the Sinkhorn loss function to bin at a lower resolution, depending on the diffractogram sparsity. For example, near-field diffractograms may contain over 100 million non-zero pixels, and comparisons of that size become prohibitively expensive.

A visual example of spot binning is shown in \cref{fig:spot_binning}. The position of a spot, in general, does not lie in the center of the binned pixels; in this sense, there is some notion of resolution below the bin width.

\subsubsection{Initial guess}\label{sec:initial_guess}

Given that the optimization problem in \cref{eq:general_inverse_problem} is non-convex, one must supply an initial guess near the minima of interest. Standard HEDM reconstruction algorithms provide a good starting point where either i) a near-field reconstruction or ii) far-field tessellation approximation can be used as an initial guess. In this work, we generate the initial guess by a standard near-field Monte Carlo orientation search over some set of trial orientations (generated by far-field seeding from a nearest-neighbor search). The methods presented here provide no mechanism to add grains that were missed in the initial guess, so the trial orientations should be dense enough that small grains simply exist in the initial state. The initial guess for the traction distribution is a uniform traction field on the top and bottom boundaries with net load equal to the macroscopic load at the measured state. 

\subsubsection{Overall algorithm}

\begin{figure}
    \centering
    \includegraphics[width=\textwidth]{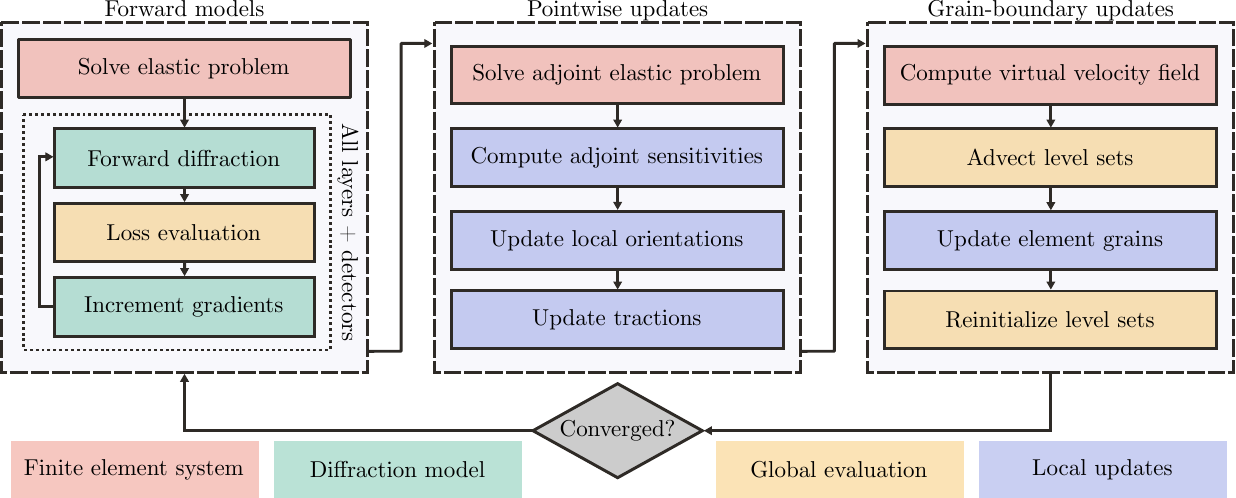}
    \caption{Broad algorithm for a single optimization iteration of the reconstruction method developed in this work. Each box is colored by the category of its performed computation.}
    \label{fig:overall_algorithm}
\end{figure}

The overall algorithm is shown in general in \cref{fig:overall_algorithm}, where each iteration is split into three stages: forward modeling to compute all necessary gradients of the objective function, pointwise updates for the orientation and traction fields, and grain boundary updates from the multiple level set approach. The algorithm presented here is implemented in shared-memory parallel C++ code, which we call {\texttt{PARA-X}}. All finite element--related operations are evaluated using the deal.II finite element library \cite{arndtDealIILibraryVersion2025} with a static unstructured tetrahedral mesh and linear (P1) Lagrange elements. Trilinos \cite{thetrilinosprojectteamTrilinosProjectWebsite} is used to solve the linear systems with a conjugate-gradient solver and algebraic multigrid preconditioner, and six degrees of freedom are constrained to remove the rigid body modes. As noted earlier, the Sinkhorn distance is computed using the GeomLoss library by \citeauthor{feydyGeometricDataAnalysis2020}~\cite{feydyGeometricDataAnalysis2020}.

\section{Verification: Synthetic examples}\label{sec:synthetic_examples}

We now present a few examples of reconstructions performed using synthetically generated data following the forward modeling approach outlined above.

\subsection{Synthetic data generation}

We consider two main test cases: a single-layer case and a multi-layer case more representative of a real experiment. We define a layer as a single near- and far-field diffractogram from some volume of the microstructure. For consistency between the synthetic examples and the experimental example provided later in \cref{sec:experimental_example}, we use similar experimental conditions. Specifically, we use aluminum oxynitride (AlON) as the test material, with material constants provided in \cref{table:alon_data}. For other parameters specific to the near- and far-field modalities, we use those given in \cref{table:parameters}. We generate a $1\,\text{mm} \times 1\,\text{mm} \times 1\,\text{mm}$ equiaxed microstructure with random texture using DREAM.3D \cite{groeberDREAM3DDigitalRepresentation2014}; note that there is no mosaicity in these synthetic microstructures (i.e., the intragranular misorientation is zero). The microstructure is meshed with 5 million elements, the top and bottom faces of the microstructure are uniformly loaded with a net $-850$\,N compressive load (corresponding to the approximate state of the experimental data in \cref{sec:experimental_example}), and the strain field is computed from a forward elastic finite element simulation. Using a $100$\,\textmu{}m tall box beam, we compute five synthetic near- and far-field diffractograms, each vertically offset by the beam height such that a net volume of $1000\,\text{\textmu{}m} \times 500\,\text{\textmu{}m} \times 1000\,\text{\textmu{}m}$ is interrogated in the center of the polycrystal. A visualization of this process is shown in \cref{fig:synthetic_data_generation} for clarity. We only compute kinematic diffraction events (single scattering), use a constant beam intensity profile, and do not include attenuation. This study is entirely ideal and is best suited for interpretability of methods presented here rather than as a detailed parametric study. 

In what follows in the verification studies, the meshes over which the fields are reconstructed are a strict subset of the mesh used to generate the data (i.e., there is an exact one-to-one comparison between elements). We also use the near-field data for only grain-boundary updates, and the far-field data are used for orientation and strain field updates. Additionally, the orientation of each point is assigned the grain-averaged value after each optimization step (i.e., we enforce zero mosaic spread as is known in the ground truth). Other update schemes are possible (e.g., see \cref{sec:experimental_example} and \cref{sec:update_modalities}).
\begin{table}
    \centering
    \caption{Constants for aluminum oxynitride (AlON) \cite{satapathySingleCrystalElasticProperties2016}. The linear attenuation coefficient $\mu$ is approximated for AlON at 65.35\,keV from the NIST XCOM dataset \cite{seltzerXCOMPhotonCrossSections1987}.}
    \label{table:alon_data}
    \begin{tabular}{cccccc}
        \toprule
        $c_{11}$ (GPa) & $c_{12}$ (GPa) & $c_{44}$ (GPa) & Attenuation coeff.\ $\mu$ (m$^{-1}$) & Lattice parameter, $a$ (\AA) & Space group \\
        \midrule
        334.8 & 164.4 & 178.6 & 82 & 7.945 & 227 \\
        \bottomrule
    \end{tabular}
\end{table}
\begin{table}
    \caption{Beam, detector, multiscale Sinkhorn, and other parameters used for both the experiment and the synthetic data. We set the blur to be at the scale of the bin width (data spacing). The point spread is visually approximated for each detector and represents the standard deviation of a 2D Gaussian PSF. Most parameters are consistent between the synthetic examples and the experimental example; those that are not are denoted as such by ``synthetic'' or ``experimental.''}
    \centering
    \label{table:parameters}
    \begin{tabular}{lcc}
        \toprule
        Parameters & Near-field & Far-field \\
        \midrule
        \multicolumn{3}{l}{\textit{Monochromated beam}} \\
        \quad Energy, $E$ (keV) & 65.35 & 65.35 \\
        \quad Bandwidth, $\Delta E / E$ & $10^{-3}$ & $10^{-3}$ \\
        \quad Synthetic height $\times$ width (mm) & $0.1 \times 3$ & $0.1 \times 3$ \\
        \quad Synthetic profile & Uniform & Uniform \\
        \quad Experimental height $\times$ width (mm) & $0.001 \times 1.8$ & $0.1 \times 2.1$ \\
        \quad Experimental profile & Parabolic & Parabolic \\
        \addlinespace
        \multicolumn{3}{l}{\textit{Detector}} \\
        \quad Model & Scintillator + Retiga 4000DC & GE 41RT \\
        \quad Distance, $L_\mathrm{sd}$ (m) & 0.0085 & 1.44 \\
        \quad Rotation range (deg) & 180 & 360 \\
        \quad Rotation step, $\Delta \omega$ (deg) & 0.25 & 0.25 \\
        \quad Pixel size (\textmu{}m) & 1.48 & 200 \\
        \quad Pixel dimensions & $2048 \times 2048$ & $2048 \times 2048$ \\
        \quad Pixel depth (bits) & 12 & 14 \\
        \quad Point spread (px) & 2 & 1 \\
        \quad Simulation alignment & Centered & Centered \\
        \quad Experimental alignment & Bottom & Centered \\
        \addlinespace
        \multicolumn{3}{l}{\textit{Multiscale Sinkhorn parameters} \cite{feydyGeometricDataAnalysis2020}} \\
        \quad Blur, $\sqrt{\epsilon}$ & $2 \times 10^{-3}$ & $5 \times 10^{-4}$ \\
        \quad Reach, $\sqrt{\rho}$ & $4 \times 10^{-3}$ & $2.5 \times 10^{-3}$ \\
        \quad Scaling & 0.9 & 0.9 \\
        \quad Binning dimensions $(x,y,\omega)$ & $512 \times 512 \times 720$ & $2048 \times 2048 \times 1440$ \\
        \addlinespace
        \multicolumn{3}{l}{\textit{Other}} \\
        \quad Velocity smoothing, $\alpha$ (\textmu{}m) & $10$ & --- \\
        \quad Experimental points per element & 20 & 4 \\
        \quad Synthetic points per element & 4 & 1 \\
        \quad Mask angle, $\eta_\mathrm{mask}$ (deg) & 5 & 10 \\
        \bottomrule
    \end{tabular}
\end{table}

\subsection{Error metrics}\label{sec:error_metrics}

Given a synthetically generated ground-truth microstructure, the exact strain and orientation fields are known, so quantitative ground-truth error metrics for the reconstruction can be computed. The pointwise and macroscopic errors in the strain field are computed as
\begin{equation}
    E(x) \coloneqq \|\varepsilon_\mathrm{curr}(x) - \varepsilon_\mathrm{gt}(x)\|_2, \quad \overline E \coloneqq \frac{\frac{1}{|\Omega|} \int_\Omega E(x) \dif x}{\|\varepsilon_\mathrm{macro}\|_2}
\end{equation}
where the strain tensors are written and computed in symmetric Mandel notation, and the subscripts indicate the current, ground truth, and macroscopic strain. The pointwise error in the orientation field is computed as the minimum misorientation angle (i.e., considering crystal symmetry) between the ground truth and the reconstructed orientation. This is distinct from intragranular misorientation, which describes mosaic spread from the mean grain orientation.

To provide a familiar measure from existing methods, we also compute the completeness $\mathcal C$ throughout. We define completeness in the standard manner, where for each point we compute the ratio of the number of simulated diffraction spots that overlap with the experimental data to the number of simulated spots. In this work, we compute completeness for spots that are bright enough to possibly appear on the detector (i.e., for simulated points that appear after the culling step in \cref{sec:binning}) and compute overlap with the filtered images at the measured resolution. We additionally compute an ``off-by-one'' completeness, which counts spots as overlapping if they appear on the simulated rotation frame, or either the previous or next frame. For the purposes of better highlighting mean completeness improvements in log space, we plot the mean ``incompleteness,'' which we define as $1 - \mathcal C$.

\subsection{Single-layer case}\label{sec:single_layer_case}

In the first numerical example, we evaluate reconstruction performance on a single $100$\,\textmu{}m scan layer with 500,000 elements. For this simple case, we apply uniform loading on the top and bottom boundaries for both the synthetic data and reconstruction and do not perform traction updates. To this end, we evaluate the reconstruction method in an orientation-focused regime. We also perform the reconstruction twice, with and without level set grain boundary updates, to highlight the importance of correctly capturing grain topology.

\begin{figure}
    \centering
    \includegraphics[width=\textwidth]{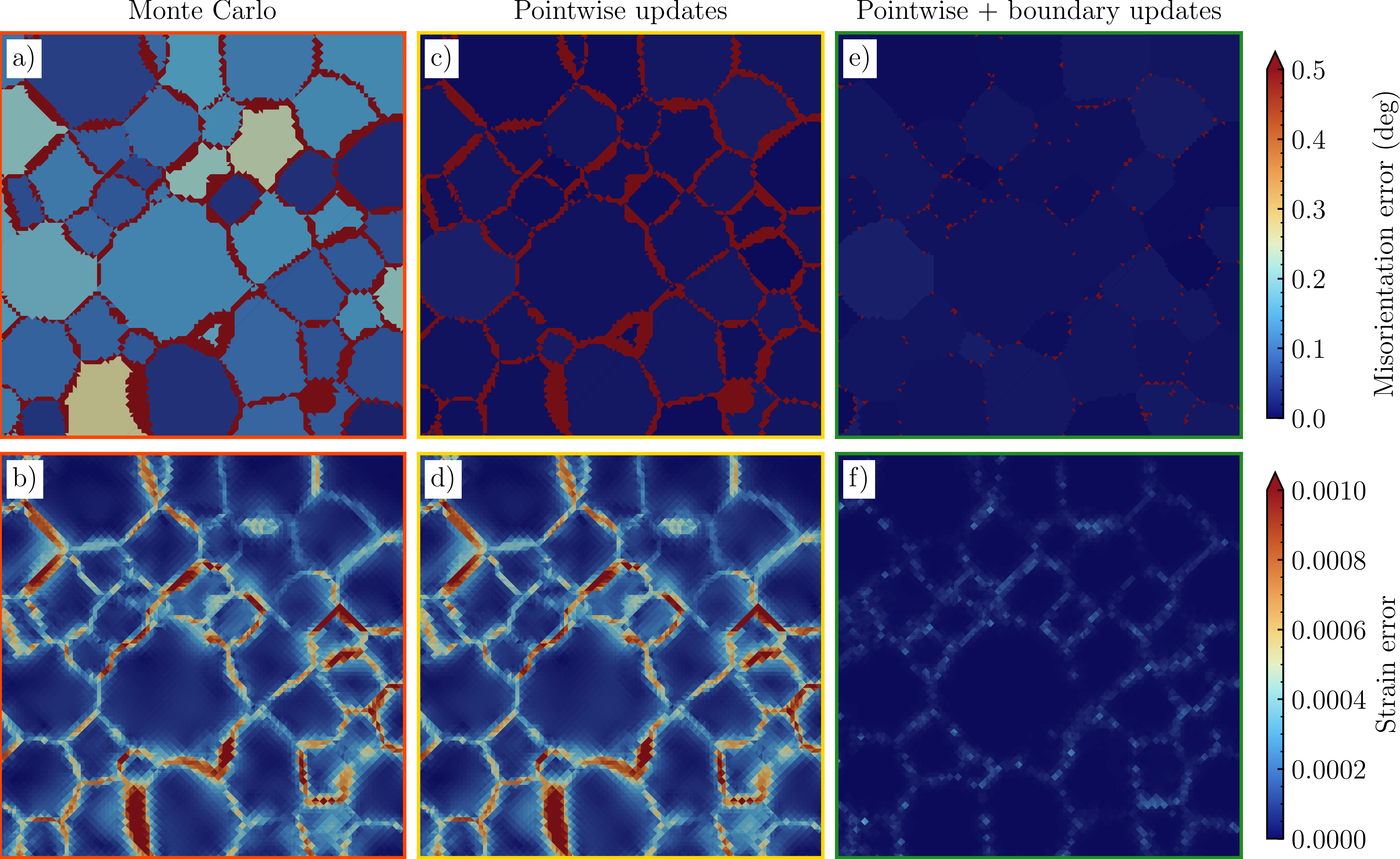}
    \caption{Pointwise misorientation and strain error between the synthetic ground-truth microstructure and the a,b) initial Monte Carlo near-field search, c,d) pointwise orientation update only, and e,f) coupled pointwise and level set grain boundary update. The fields are shown for a 2D slice in the vertical center of the layer.}
    \label{fig:monte_carlo_vs_local_vs_boundary}
\end{figure}

The misorientation and strain error between the ground truth and reconstructions on a slice through the layer are shown in \cref{fig:monte_carlo_vs_local_vs_boundary}. From \cref{fig:monte_carlo_vs_local_vs_boundary}a and b, we observe that the local updates are able to capture the orientations of the grains correctly, but are unable to overcome grain boundary positional errors stemming from the initial Monte Carlo search. As the elasticity PDE is elliptic, the error in the strain field is not confined to the misidentified elements; it is elevated everywhere. After refinement of the orientations locally in \cref{fig:monte_carlo_vs_local_vs_boundary}c and d, the orientations of the grains are captured correctly away from grain boundaries, but the strain error shows no change as the changes in stiffness tensor are small under a small $<1^\circ$ rotation, particularly for the cubic system here. Upon the addition of level set grain boundary updates in \cref{fig:monte_carlo_vs_local_vs_boundary}e and f, the grain structure is captured significantly better than the other two cases, and this drastically decreases the local strain error everywhere. This highlights that, for a PDE-constrained problem, the reconstructed grain boundary topology must be highly accurate for strain field recovery purposes. The Monte Carlo initial guess captures the existence of most grains, but mispredicts grain boundaries initially, and this is particularly accentuated here due to the use of a box beam for the near-field data.

\begin{figure}
    \centering
    \includegraphics[width=\textwidth]{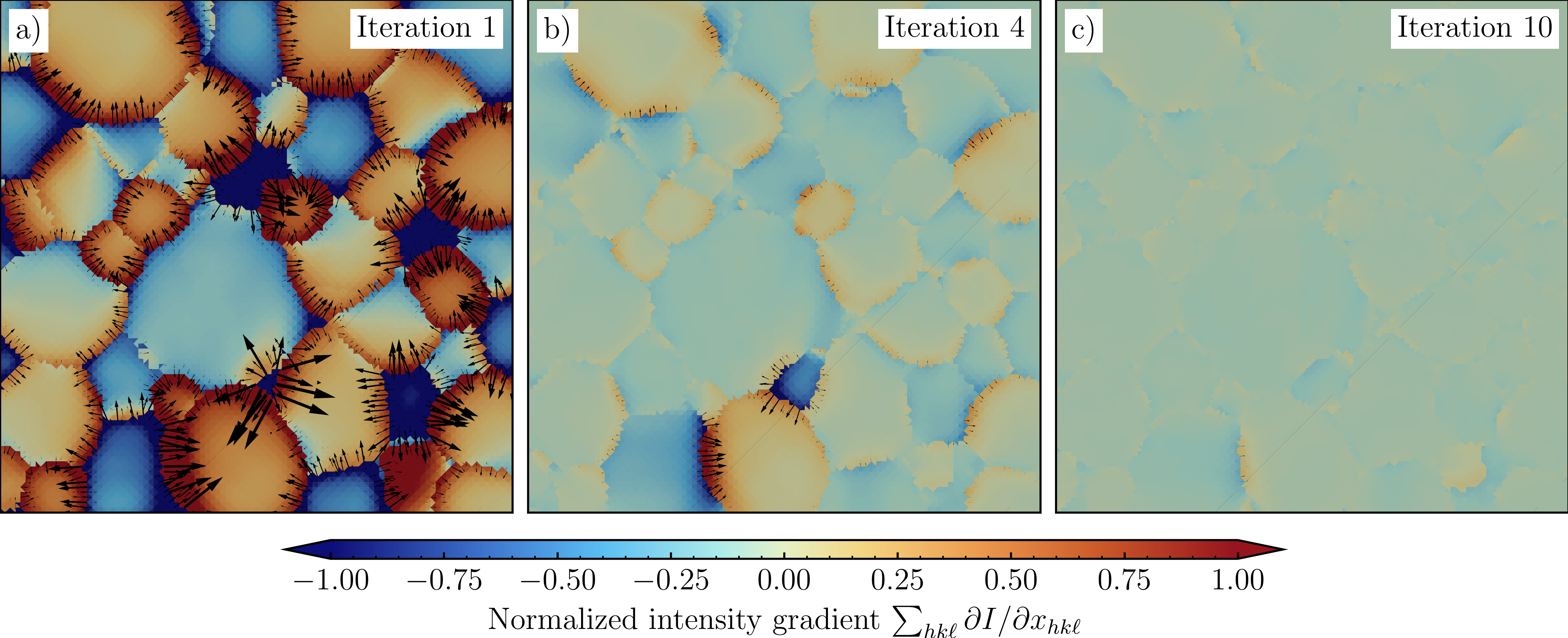}
    \caption{Grain boundary advection at the a) first, b) fourth, and c) tenth optimization iterations, where the normalized intensity gradient is shown. For each point, the intensity gradient is the summation of $\partial d / \partial I_\hkl$ over all measured reflections; this is the dominating term in \cref{eq:final_inner_variation} used to compute the virtual velocity field. The computed virtual velocity field from \cref{eq:helmholtz_velocity_field} is shown, without the out-of-plane component, by the black arrows (scaled by the velocity magnitude), which is used to update the grain topology through the multiple level set advection scheme in \cref{sec:grain_boundary_optimization}.}
    \label{fig:level_set_advection_layer}
\end{figure}

We may also examine the grain boundary advection process through the optimization iterations. \Cref{fig:level_set_advection_layer} shows the pointwise intensity gradient $\sum_{\hkl} \partial d / \partial I_\hkl$, which is the dominating term from the inner variation calculation in \cref{eq:final_inner_variation}, along with the computed virtual velocities from \cref{eq:helmholtz_velocity_field} at multiple optimization iterations. Grains that are incorrectly captured have larger velocity magnitudes due to the higher diffraction mismatch. Further, the use of spatially sensitive near-field data for grain boundary updates provides information about \emph{where} grains need to locally grow or shrink, which is not the case for far-field data. In this synthetic case, once the grain boundaries are updated to within one element of the ground truth, the velocity field has small-amplitude oscillations due to elements at grain boundaries alternating between the two grains (i.e., the level set has sub-element grain boundary interpolation, but the forward diffraction model does not). It is in this sense that the geometric uncertainty of the method as implemented is at least as large as the element size.

\subsection{Multi-layer case}\label{sec:synthetic_multilayer_example}

\begin{figure}[h]
    \centering
    \includegraphics[width=\textwidth]{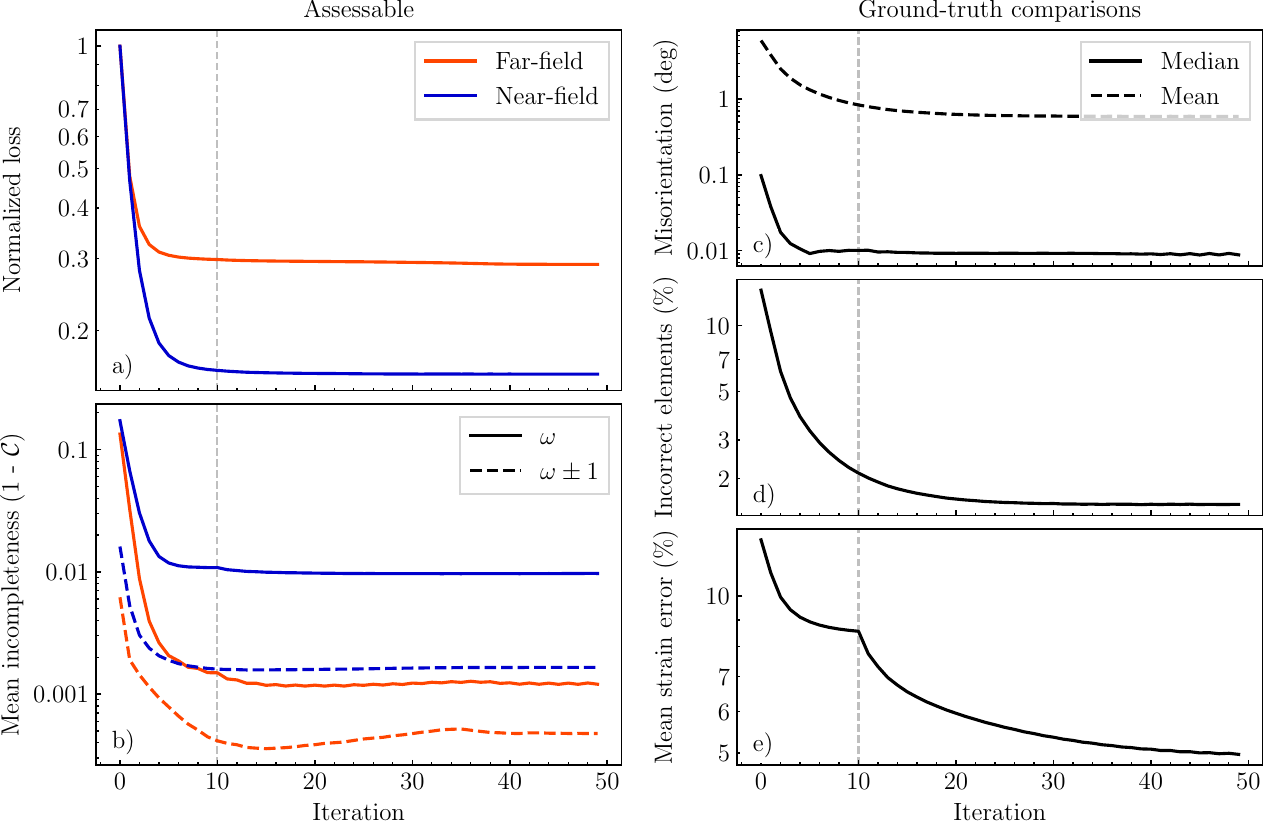}
    \caption{Optimization evolution of the a) loss function and b) mean near- and far-field (in)completenesses, which are always measurable for experiments, along with the ground-truth error metrics of c) median misorientation, d) fraction of elements with incorrect grain (i.e., elements with greater than $5^\circ$ misorientation tolerance from the ground truth), and e) mean strain error. The dashed vertical line indicates the iteration where traction updates began, and the $\omega \pm 1$ indicates the off-by-one completeness discussed in \cref{sec:error_metrics}.}
    \label{fig:loss_evolution}
\end{figure}

We now perform reconstruction over the five-layer synthetic dataset with 2.5 million elements using the full reconstruction approach outlined in \cref{sec:methods}. The evolution of the loss function, completeness, misorientation, and strain error over the optimization iterations is shown in \cref{fig:loss_evolution}. Additionally, the initial iterations fix the applied tractions to be uniform, and optimization over the traction distribution is only enabled on the tenth iteration. This results in an accelerated decrease in the total strain field error, but the other quantities are relatively unaffected as the diffractograms are less sensitive to strain than other quantites.

\begin{figure}
    \centering
    \includegraphics[width=\textwidth]{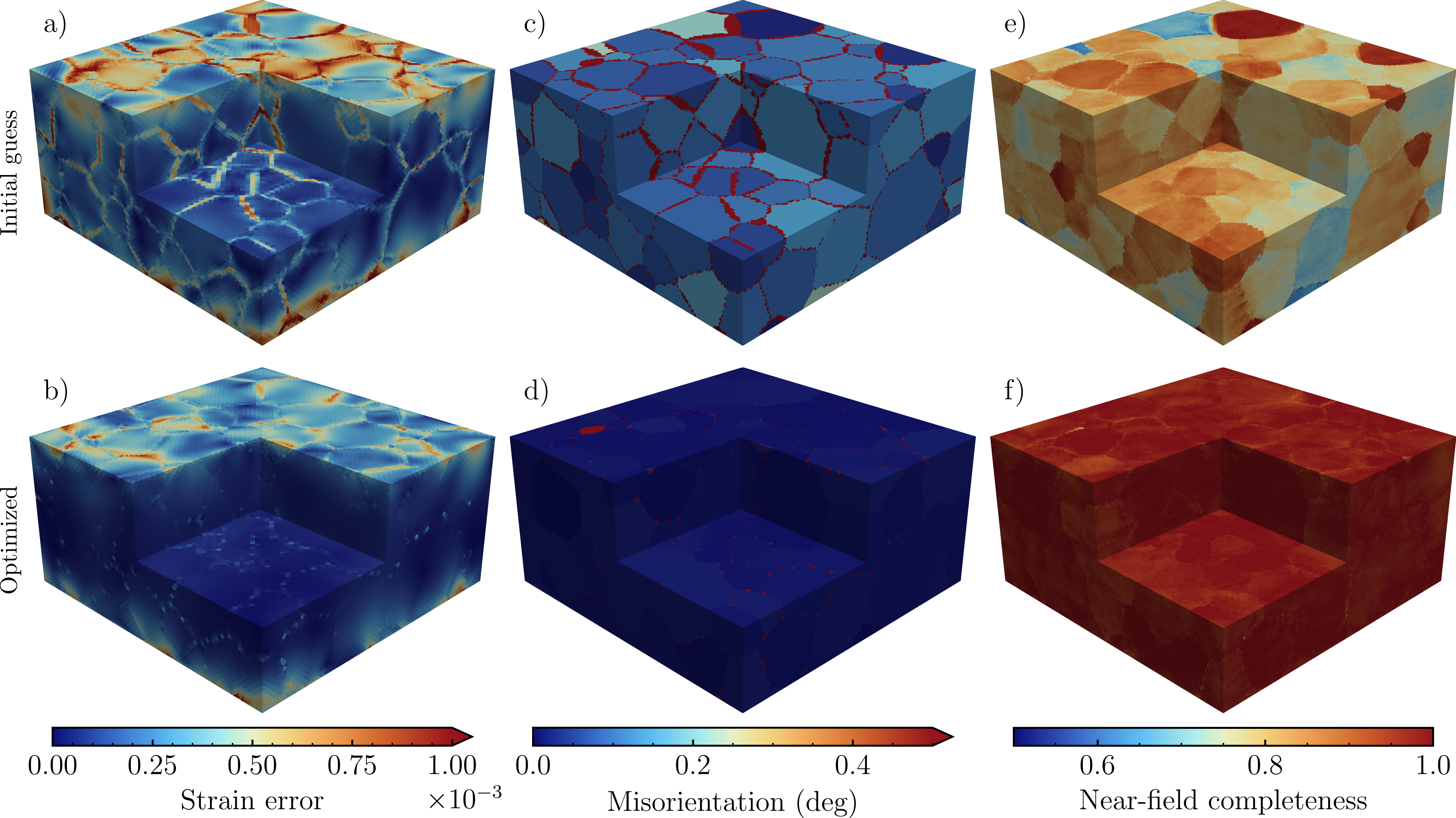}
    \caption{Initial error metrics (top row) versus final error metrics (bottom row) for the multi-layer case, showing a,b) ground-truth strain error, c,d) ground-truth misorientation, and e,f) near-field completeness.}
    \label{fig:multilayer_error_evolution}
\end{figure}

While the averaged error metrics in \cref{fig:loss_evolution} are useful, they do not provide any local information. To this end, in \cref{fig:multilayer_error_evolution} we plot the strain error, misorientation, and near-field completeness between the state of the Monte Carlo initial guess and the final optimization fields. The decrease in global error (increase in completeness) generally results from a decrease in error everywhere, though with some exceptions. First, observe that there are some grains that are not captured in the initial guess; these errors are never corrected by the reconstruction method here (cf.\ \cref{fig:multilayer_error_evolution}c and d) as there is no mechanism by which grains can be re-added to the domain during the optimization process. Also note by comparing \cref{fig:multilayer_error_evolution}c and d that the initial orientation guess can be misoriented from the ground truth (in this example by $0.25^\circ$) and is recovered in this example. In this regard, the most important requirement of an initial guess is that grains simply exist, but they may be misoriented or malformed.

It is also seen from \cref{fig:multilayer_error_evolution}a and b that the strain error at the final state is predominantly concentrated at the loaded boundaries. This follows from the same Saint-Venant arguments by \citeauthor{cockeRecoveringIntragranularStrain2025}~\cite{cockeRecoveringIntragranularStrain2025}, and reaffirms that experiments should be conducted on taller domains such that the precise local effects of the boundary conditions are negligible in the actual region of interest. Even so, the traction optimization still reduces the error at the boundaries when compared to the initial state.

Overall, this example shows that the proposed approach accurately reconstructs the synthetic data. The results in \cref{fig:loss_evolution} also demonstrate a feature that is important in analyzing experimental data. In this example, we have the ground-truth orientation, grain structure, and strain information in detail, but these are not available during a real experiment. However, the loss function and completeness can always be assessed in an experimental setting. \Cref{fig:loss_evolution} shows that the evolution of the assessable quantities---loss and completeness---is correlated with the accuracy of physically meaningful quantities like orientation, strain, and grain geometry. Therefore, we can use the former as a proxy for the latter in real experiments, noting that these proxies are less sensitive for inferring strain error.

\subsection{Computational cost}

\begin{figure}
    \centering
    \includegraphics[width=\textwidth]{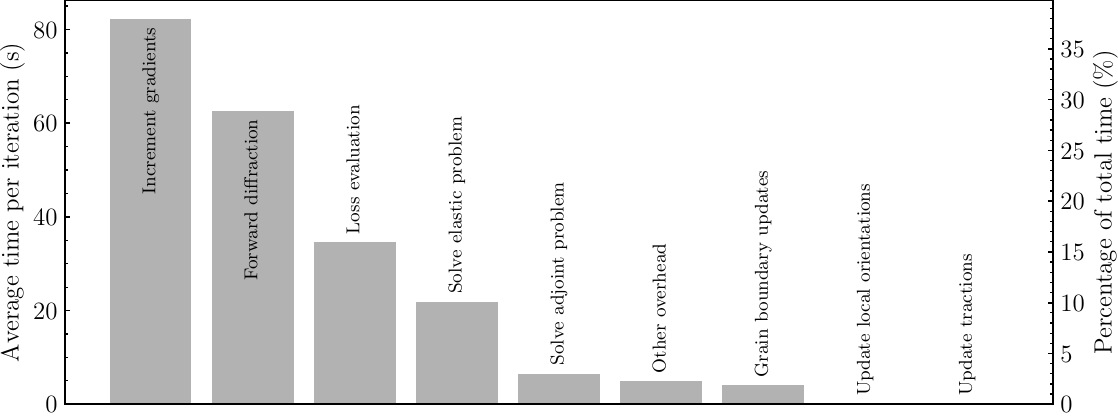}
    \caption{Computational cost breakdown of a single iteration within the optimization procedure, with each value given as the mean over all iterations. A significant portion of the forward diffraction and gradient increment runtime is associated with probing the hash maps that store the sparse diffractograms.}
    \label{fig:time_breakdown}
\end{figure}

We now show the computational cost of an average iteration from the synthetic example in \cref{sec:synthetic_multilayer_example}. All operations are computed in shared-memory parallel using 32 physical cores of an AMD EPYC 9554 CPU with an NVIDIA L40S GPU used to compute the Sinkhorn loss. The computational cost breakdown of each optimization iteration is shown in \cref{fig:time_breakdown} and the total runtime for 50 optimization iterations is approximately 3 hours. Choices for parameters within the forward model and Sinkhorn loss evaluation can significantly increase or decrease the runtime. During Sinkhorn loss evaluations, in each point cloud, the near-field comparisons contained approximately 6 million points, and the far-field comparisons contained approximately 150,000 points (i.e., the loss evaluation is entirely dominated by the near-field data, hence the more aggressive binning in \cref{table:parameters}).
One-time computational costs associated with data loading and initial finite-element-related setup are neglected here. There is also the non-negligible cost of computing the initial guess via Monte Carlo orientation search; this cost is at worst the same as that of previously developed methods \cite{suterForwardModelingMethod2006}, and can usually be smaller since it can be performed with a coarser orientation search.

\section{Validation: Experimental example}\label{sec:experimental_example}

We evaluate the behavior of the reconstruction method using real HEDM data collected on aluminum oxynitride (AlON, Al$_{23-X/3}$O$_{27+X}$N$_{5-X}$) under conditions similar to those in the synthetic examples in \cref{sec:synthetic_examples}. These data have been used in a study of brittle fracture by \citeauthor{gorskeInsituVisualizationGrowing2026a}~\cite{gorskeInsituVisualizationGrowing2026a} (sample 1), and are also the data on which the equilibrium-constrained post-processing method by \citeauthor{cockeRecoveringIntragranularStrain2025}~\cite{cockeRecoveringIntragranularStrain2025} was tested. 

\subsection{Experimental details}\label{sec:experimental_details}

We briefly outline the experimental details below, though further details are provided by \citeauthor{gorskeInsituVisualizationGrowing2026a}~\cite{gorskeInsituVisualizationGrowing2026a}, and most key experimental details are provided in \cref{table:parameters}. The AlON sample was fabricated as a compression parallelepiped with design dimensions of $1.4\,\text{mm} \times 16\,\text{mm} \times 1.4\,\text{mm}$ and drilled with a hole such that stable fracture would occur during compression. The sample was interrogated with near- and far-field HEDM at the 1-ID beamline at the Advanced Photon Source at Argonne National Laboratory. The beam was first monochromated to the Hf K-edge of 65.35\,keV and the sample was placed in compression platens within the RAMS-III load frame \cite{shadeRotationalAxialMotion2015}. The sample was then loaded and held at multiple measurement states. At the state we use for the experimental analysis here, the sample was compressed to $-1700$\,N, corresponding to an approximate engineering stress and strain of $-850$\,MPa and $-0.26$\%, respectively. In total, an uncracked volume of approximately $1370\,\text{\textmu{}m} \times 300\,\text{\textmu{}m} \times 1460\,\text{\textmu{}m}$ was interrogated by overlapping near- and far-field scans, with 61 total near-field layers vertically spaced $5$\,\textmu{}m apart and three total far-field layers vertically spaced $100$\,\textmu{}m apart. The sample was held in load control for the duration of these measurements, after an initial $50$\,N unload. Near-field measurements were taken at two detector distances, though we only use data from the closer of the two. The total duration of these scans was approximately 19 hours, during which it was assumed the only change in the experimental state is the vertical sample position (no creep, sample slip or tilt, rotation drift, beam changes, etc.). 

After raw data collection, the diffraction images are filtered to extract the sparse signal. For both near- and far-field images, we first apply a subtractive temporal median filter to remove any time-persistent pixel data. For the near-field images, we perform a $3 \times 3$ pixel spatial median filter to remove ``zingers'' (rogue single-pixel peaks). For the far-field images, single non-zero pixels surrounded by all zero values are removed; this removes some zingers without disrupting the intensity distributions of the real signal. Filtering changes the total intensity of a diffraction spot, so these choices are deliberate to maximize noise reduction and sparsity while retaining signal. Once filtered, for each image, all pixels above a set threshold value are inserted into the sparse spot data structure and used as the reference dataset. We use a threshold value of 15 and 10 a.u.\ for the far-field and near-field images, respectively. Detector geometries are calibrated with a gold bi-crystal sample centered on the rotation stage, using the same optimal transport objective developed in this work. Further geometric refinement is performed during the optimization (i.e., the polycrystal fields and detector geometries are simultaneously optimized). Note that the macroscopic load constraint from \cref{eq:self_equilibrium} prevents this simultaneous geometry optimization from introducing a spurious hydrostatic offset.

Given reconstructed computed tomography (CT) measurements \cite{gorskeInsituVisualizationGrowing2026a}, the domain was approximated with flat faces and meshed with 3.35 million tetrahedral elements. The beam during the experiment exhibited a strong parabolic profile as measured from bright-field images collected during CT measurements. For all forward diffraction calculations, the intensity profile of the beam was approximated by a 1D $x_1$-varying distribution by vertically averaging over the full box beam. This is likely a good assumption for the far-field box beam but not the near-field beam as it undergoes line-focusing. We hypothesize the lack of precise spatial and temporal intensity measurements of the individual near- and far-field beams is the largest source of error in the predicted intensities of the forward model here. The width of the near-field beam was also smaller than the maximum sample dimension, so some corners of the sample leave the beam during the rotation; this is accounted for in the forward diffraction model. Following the methodology presented in \cref{sec:methods}, we compute an initial guess following standard Monte Carlo completeness maximization from the near-field data. To ensure the Monte Carlo search is not penalized by poor calibration, the search is performed with highly optimized detector geometry from a preliminary solve of the present method. The baseline Monte Carlo comparison then represents a best-case scenario. We then average the orientations over each grain (a total of 125 grains), which reduces the initial completeness due to removal of mosaicity, but improves the final optimized state.

\subsection{Reconstruction}

We now outline key details used during the reconstruction, with the main parameters given in \cref{table:parameters}. Only near-field data are used for the updates as the far-field data were found to have a varying (approximately $0.1^\circ$) rotation offset from the near-field data, so we choose not to simultaneously use both modalities like in \cref{sec:synthetic_examples}. We do, however, still compute the far-field diffractograms to assess the completeness and distance. Orientations are averaged over each grain for the first five iterations, and the traction distribution is only updated after the first ten iterations. This ensures that the tractions are only updated once grains are close enough to their final topology. The near-field beam is line-focused with a total thickness less than the vertical layer spacing; to illuminate all elements, we take the beam to be 5\,\textmu{}m tall and project the $x_2$-position of points within the illuminated volume to the beam center. We use the full intensity calculation outlined in \cref{sec:intensity}, neglecting temporal intensity changes, which may be important.

\begin{figure}
    \centering
    \includegraphics[width=\textwidth]{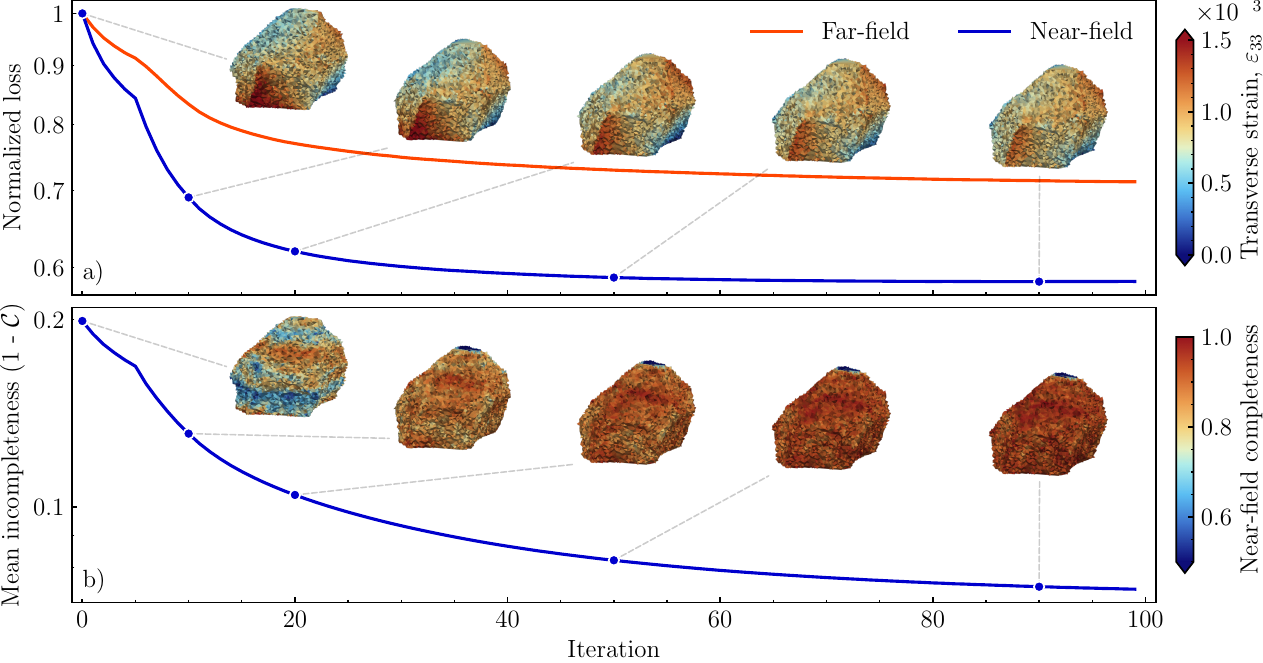}
    \caption{a) Evolution of the near- and far-field loss during the optimization, with the transverse strain component of a grain contained in all layers. b) Evolution of the mean near-field (in)completeness in the domain, with the completeness evolution of the same grain shown. Recall that \emph{only} the near-field loss is directly minimized, so the far-field loss is only shown for evaluation purposes (it is not used to compute any gradients).}
    \label{fig:experimental_loss_evolution}
\end{figure}

The methods developed in this work are now used to reconstruct the full-field state of the AlON polycrystal from the experimental data. The total runtime for 100 gradient-descent iterations is 21 hours, or approximately 13 minutes per iteration. The cost per iteration is higher for the experimental data than the synthetic data in \cref{sec:synthetic_multilayer_example} due to attenuation calculations, more elements with higher element subsampling, and more near-field layers due to the line-focused beam. A total of 114 grains are captured in the final optimized solution (i.e., 11 generally small grains are removed during the grain boundary updates). In \cref{fig:experimental_loss_evolution} we plot the near- and far-field loss and near-field completeness evolution during the optimization, along with five snapshots of a grain as its intragranular fields and shape evolve through the optimization. We do not plot the far-field completeness as it generally remains constant ($\approx 0.95$) during the optimization. From \cref{fig:experimental_loss_evolution}a, the near-field loss, which is directly optimized, monotonically decreases as is expected from a standard optimization procedure. Further, the far-field loss, which does not contribute to any gradient calculations, also monotonically decreases, highlighting that the minimization is not simply overfitting to the near-field data. These decreases in loss also directly correlate with a monotonic increase in completeness as seen in \cref{fig:experimental_loss_evolution}b. Recall from \cref{sec:synthetic_examples} that these are meaningful proxies for actual ground-truth error. The evolution of transverse strain and near-field completeness in the example grain in \cref{fig:experimental_loss_evolution} also highlights that the fields and grain topology change significantly during the optimization.

\begin{figure}
    \centering
    \includegraphics[width=\textwidth]{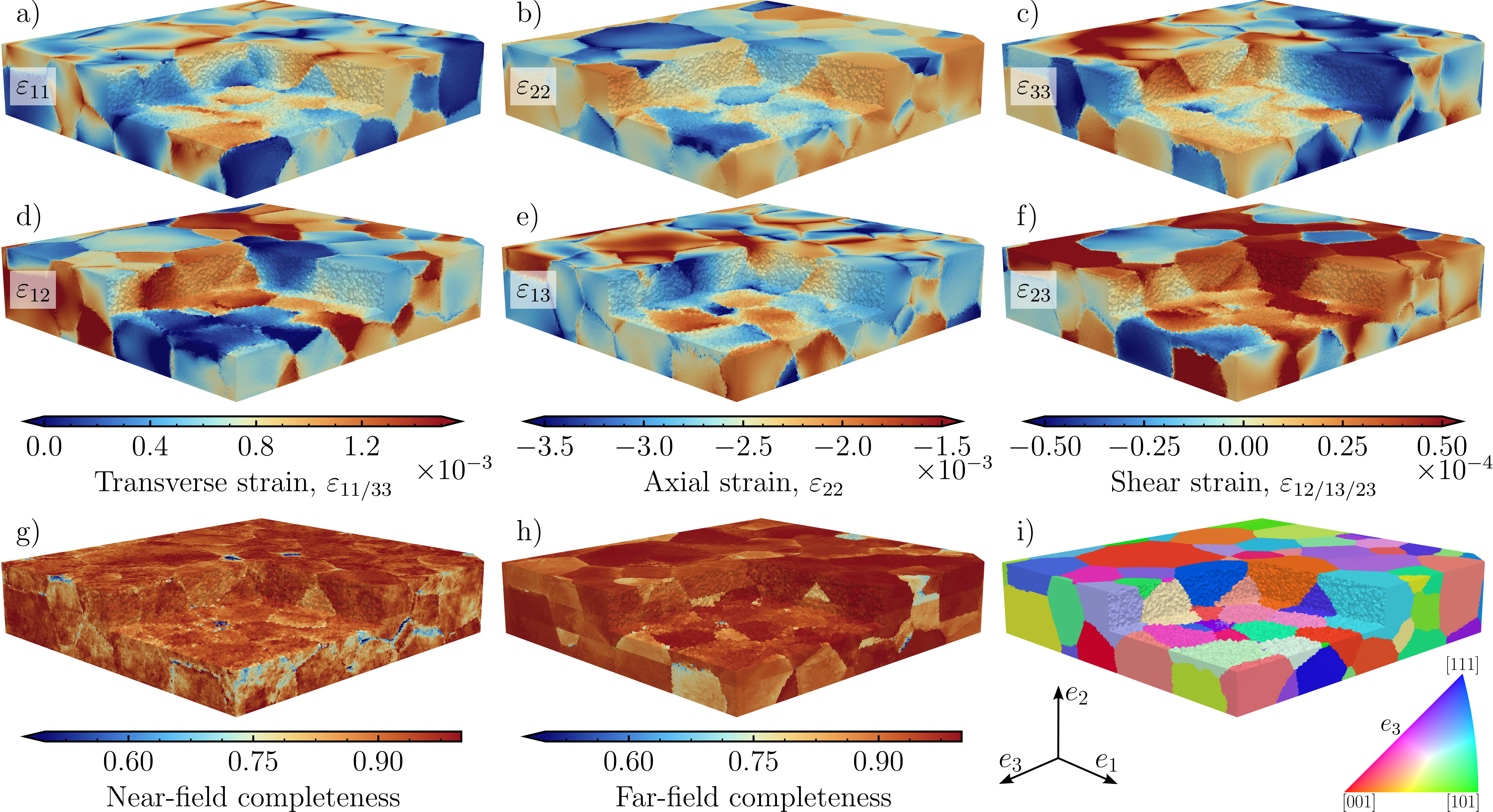}
    \caption{Final optimized strain fields plotting the a) $\varepsilon_{11}$, b) $\varepsilon_{22}$, c) $\varepsilon_{33}$, d) $\varepsilon_{12}$, e) $\varepsilon_{13}$, f) $\varepsilon_{23}$ components, and final g) near-field completeness, h) far-field completeness, and i) grain structure with inverse pole figure (IPF) coloring. Green regions in the far-field completeness are those in which no spots exceeded the threshold for visibility on the detector, which occurs when only a small portion is illuminated by the beam.}
    \label{fig:experimental_optimized_fields}
\end{figure}

Moving to the final optimized state, in \cref{fig:experimental_optimized_fields}, we show the strain field, completeness maps, and the grain structure from the entire illuminated portion of the polycrystal. From \cref{fig:experimental_optimized_fields}a--f, we observe that the recovered strain fields have large local fluctuations, especially near grain boundaries. These intragranular strain fields cannot be captured by standard reconstruction methods. The fields recovered here are not only impossible to resolve with standard methods, but are also mechanically admissible and consistent with the measured diffraction data. From the near-field completeness map in \cref{fig:experimental_optimized_fields}g, the completeness is nearly perfect (close to a value of 1) in the domain, other than some regions near the domain exterior or grain boundaries. These low-completeness areas are mostly due to missing grains in the initial guess.

The methods we follow in this work were developed after the experiment was conducted, so experimental choices are \emph{not} ideal from the standpoint of method validation. Limitations include that: i) AlON has general compositional fluctuations dependent on the oxygen content \cite{mccauleyAlONBriefHistory2009}, so any uncertainty in the undeformed crystal structure compounds with uncertainty due to experimental non-idealities; ii) only near-field data are used in the optimization due to the small drift between the near- and far-field modalities; and iii) the near-field beam intensity was not accurately resolved. These non-idealities limit insights, so future studies ought to be performed with ideal materials and methods and compared with existing experimental methods (e.g., comparisons with point-focused data sets \cite{henningssonIntragranularStrainEstimation2021} or electron backscatter diffraction serial sectioning data \cite{sparks3DReconstructionHighEnergy2024}). Regardless, considering all of these limitations, we still see significant improvement across all quantitative and qualitative near-field metrics when compared to the current de facto standard Monte Carlo optimization. 

\begin{figure}
    \centering
    \includegraphics[width=\textwidth]{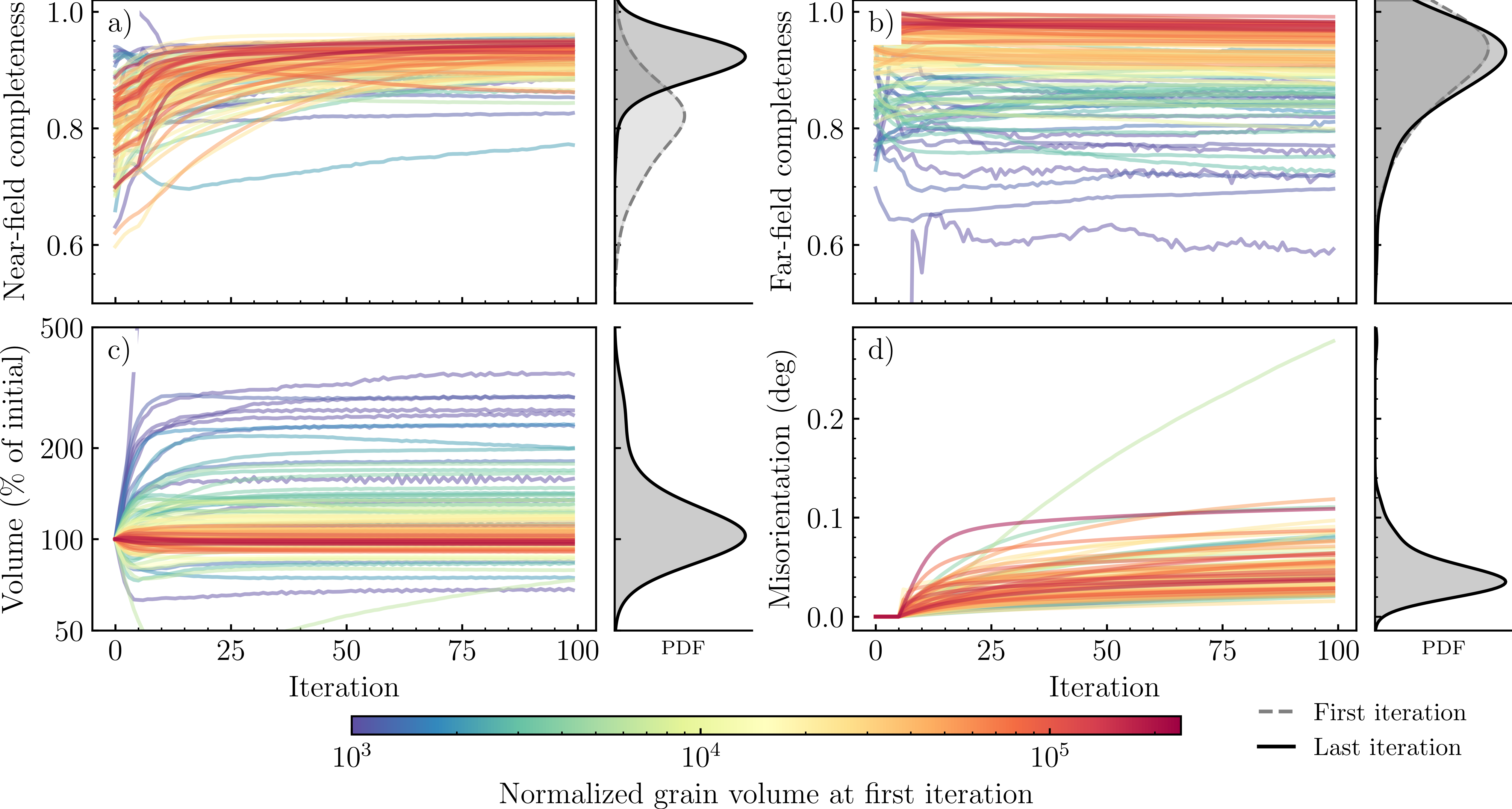}
    \caption{Plots showing, for each grain, the evolution over the optimization for the grain-averaged a) near-field and b) far-field completeness, c) volume change from the initial guess, and d) intragranular misorientation. The color of each grain is determined by the initial grain volume to better highlight how grains of different sizes evolve from the initial Monte Carlo near-field guess. Smoothed histograms are plotted for distributions of each parameter at the first and last iteration using a Gaussian kernel density estimate.}
    \label{fig:grain_data_evolution}
\end{figure}

It is also instructive to examine the individual grain-level quantities as they evolve over the optimization procedure. In \cref{fig:grain_data_evolution}, the grain-averaged values of completeness and intragranular misorientation are plotted, along with the volume change of each grain, with coloring denoting the initial grain size. The largest grains are not completely captured by the scans, so ``grain-averaged'' should be taken as ``grain scan volume-averaged.'' From \cref{fig:grain_data_evolution}a, we see that the near-field completeness starts higher than that of larger grains---this is an artifact of the initial grain-averaging. However, in \cref{fig:grain_data_evolution}c, the grains that grow the most are those with the smallest initial volume. The Monte Carlo search has much more uncertainty near grain boundaries, so small grains inherently have more topological error in these methods due to their larger surface-area-to-volume ratio. The grain-boundary optimization approach developed here corrects these errors, generally growing the smaller grains to their true size. From \cref{fig:grain_data_evolution}b and d, we note no obvious trends for the far-field completeness, which remains constant for most grains, or for the misorientation distributions in grains of different sizes. 

\subsection{Forward diffraction comparison}

\begin{figure}[h]
    \centering
    \includegraphics[width=\textwidth]{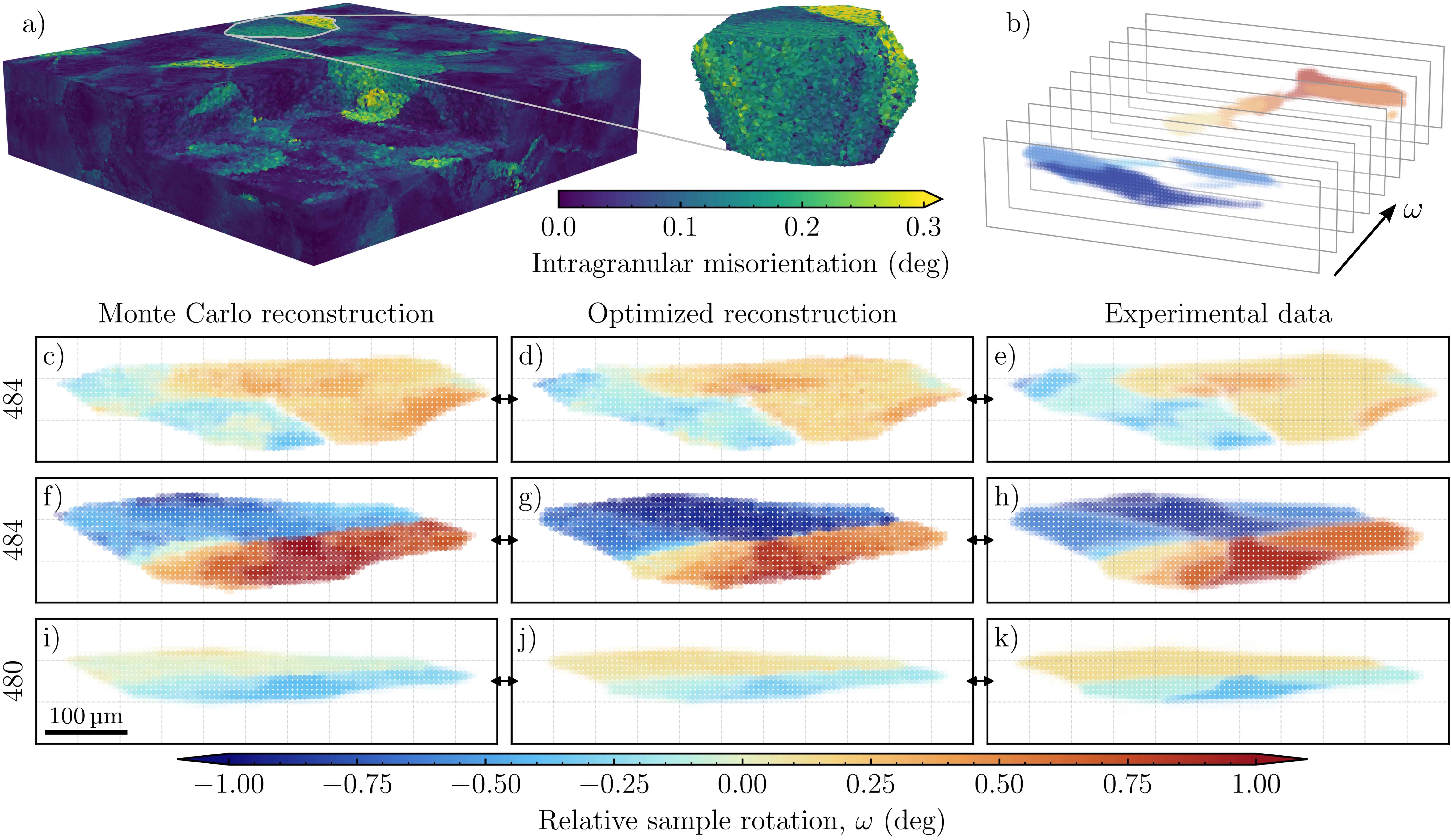}
    \caption{Comparison of single diffraction spots between the forward and the experimentally measured diffractograms. a) The intragranular misorientation of the optimized microstructure, with a highly mosaic grain isolated. b) Example of a spot diffraction signal spread over multiple rotation frames. For the highly mosaic grain shown in a), simulated diffraction spots from the Monte Carlo and optimized reconstructions are compared with the experimental measurements for three different reflections, respectively: c--e) the first $484$ reflection, f--h) the second $484$ reflection, and i--k) a $480$ reflection, where the color indicates the rotation of the spot (i.e., the third dimension of a diffractogram). Point opacity is linear with respect to the measured intensity, and the scale and axis limits for each comparison are the same.}
    \label{fig:grain_mosaic_spread}
\end{figure}

We now compare the resultant forward diffractograms at the final optimized state to the experimental diffractograms. Because diffractogram agreement is precisely the objective of the optimization procedure, these forward model comparisons are not independent measures of validation. Instead, qualitative diffractogram agreement demonstrates whether the Sinkhorn loss is meaningful as a metric. We provide a full diffractogram comparison in the supplementary material, and here instead focus on diffraction spots associated with a single highly mosaic grain within the microstructure. In \cref{fig:grain_mosaic_spread}, from the indicated highly mosaic grain, we plot three simulated spots from the Monte Carlo search (before averaging orientations over the grains) and the final optimized state and compare with the experimental measurement. It can be seen from the measured diffraction spots in \cref{fig:grain_mosaic_spread}e, h, and k that the grain has at least one strong internal low-angle grain boundary as attested by the bifurcated diffraction signal; this low-angle grain boundary is recovered in the reconstruction (see \cref{fig:grain_mosaic_spread}a). Further, by comparing \cref{fig:grain_mosaic_spread}c--e, f--h, and i--k, the forward-modeled substructure is better captured by our method than the Monte Carlo search, both in total shape and intra-spot rotational agreement. Quantitatively, the optimized reconstruction outperforms the (non-averaged) Monte Carlo reconstruction: the mean completeness increases from $0.884$ to $0.926$ and the objective decreases from $7.82 \times 10^{-4}$ to $5.48 \times 10^{-4}$. Remark also that the optimization initially averages the orientations over each grain, so the intragranular misorientation distribution is captured entirely through the gradient procedure detailed in this work rather than anything induced by the Monte Carlo search. This again highlights that the initial guess does not have to be perfect for this optimization procedure.

\section{Concluding remarks}\label{sec:discussion}

We have developed a method to reconstruct the grain structure and full strain fields of elastic polycrystals using HEDM data. The main points of departure from existing methods are as follows:
\begin{itemize}
    \item The complete diffractograms obtained from experiment and the forward model are compared globally using a differentiable distance. This enables a gradient-based optimization over the entire polycrystal and eliminates the need to solve per-grain sub-problems.
    \item The method incorporates measured intensities directly in the objective and does not require experimental data labeling (viz.\ indexing), allowing for reconstruction refinement from diffraction spot substructure, even when spots may overlap.
    \item The physics of deformation is directly incorporated into the forward problem, so the resultant fields are physically meaningful.
    \item Grain boundary topology is refined using the measured diffractograms, and grain-to-grain interactions are explicitly considered due to the global nature of the reconstruction.
    \item The framework is fully compatible with preexisting datasets collected for traditional HEDM reconstruction and works with standard near- and far-field detectors and experimental setups.
\end{itemize}
During verification of the method using synthetic data, the microstructure was reconstructed by minimizing the optimal transport distance, which improved all ground-truth error metrics (misorientation, strain error, and grain structure). Reconstructions performed on real aluminum oxynitride near-field HEDM data showed a strong increase in near-field completeness and identified intragranular orientation structure with an initially zero-mosaicity guess. Imprecise alignment between near- and far-field modalities precluded further full-field analysis, which remains an open challenge for simultaneous use of multiple diffraction modalities. As the methodology and corresponding experiments are refined, HEDM moves from being an observation tool toward a tool by which one can learn detailed polycrystalline physics.
\vspace{\baselineskip}

We conclude with comments on the continued development of this method.

\paragraph{Computation}

A main motivating factor for reconstruction algorithms is to move toward real-time in situ reconstructions. During experiments conducted at synchrotron beamline facilities, users are often limited to experimental data collection windows of only a few days. If the framework presented here can be evaluated in near real-time, the resultant reconstruction may be used to inform and drive the experimental procedure. The gradient-based methods presented here are also well suited for this, in that fields can be quickly updated from a previous state if the microstructural changes are small. The computational cost of the method developed here, at around 4 minutes (5 layers) and 13 minutes (61 layers) per iteration for the synthetic and experimental cases on a single shared-memory system, is comparable to, or lower than, the cost of existing methods for recovering intragranular strain \cite{shenVoxelbasedStrainTensors2020,reischigThreedimensionalReconstructionIntragranular2020}. Even so, there are numerous directions in which the framework presented here may be modified to reduce computational costs. For parametric updates, conjugate-gradient or quasi-Newton methods may provide more rapid convergence, and the cost of updates is negligible compared to the forward model and loss evaluations (see \cref{fig:time_breakdown}). Finally, the methods here are only implemented with shared-memory computing, so distributed memory or GPU computing, alongside complementary algorithm development, may be useful for performance improvements.

\paragraph{Modeling choices}

The methods developed here introduce many modeling choices and parameters that may be determined somewhat arbitrarily. For modeling choices, the main decisions involve which parameters to optimize and in what order, primarily considering that gradients may be computed with respect to most parameters within the forward model. For parameter choices, a non-exhaustive list includes parameters for experimental image filtering, forward diffraction, loss-function evaluation, optimization, and constitutive modeling. Within the powder X-ray diffraction Rietveld analysis community, a multitude of heuristics and practical guidelines have been developed for improving structure calculations \cite{mccuskerRietveldRefinementGuidelines1999}. The idea of the framework presented here follows as a 3D spatial generalization of Rietveld analysis, so it is perhaps possible to establish guidelines that lead to better reconstructions; however, the varied setup, drastically larger datasets, and geometrically richer design of HEDM experiments make designing a general-purpose set of guidelines much more difficult. At the very least, future work should aim to parametrically study this method in detail to establish core principles.

\paragraph{Forward modeling physics}

Given that we use a full forward modeling approach for both diffraction and deformation physics, reconstruction accuracy is bounded by forward model accuracy. The diffraction model used in this work is relatively simple: ideal crystals, elastic scattering, and a perfectly collimated, monochromatic incident beam are assumed. From the perspective of this framework, the \emph{only} strict requirements of any forward diffraction model are that diffraction be local and piecewise differentiable with respect to the underlying fields in the polycrystal. In the same vein, this work assumes the simplest mechanical constraint of small strain linear elasticity. X-ray studies are often performed well into the large deformation plastic regime, so future extensions ought to incorporate these additional mechanics. Further, for larger deformations, the diffraction consequences of dislocations, lattice distortions, and other physical phenomena become stronger and may need to be considered.

\paragraph{Implications for future experiments}

As we use fundamentally different ideas for reconstruction compared to existing methods, there are several considerations to make when performing experiments for a full forward modeling reconstruction approach. From an experimental standpoint, while reducing uncertainties in geometric parameters is standard (e.g., the use of calibrants), intensity-controlling factors are commonly less considered. As remarked by \citeauthor{shenVoxelbasedStrainTensors2020}~\cite{shenVoxelbasedStrainTensors2020}, the incident beam has time-dependent flux and energy distributions that are not normally accounted for during canonical HEDM reconstructions; however, ion-chamber, foil absorption edge, and beam-profile measurements are often taken during experiments, which can be used to account for these temporal changes. For the method developed in this work, direct measurement of the beam intensity profile (e.g., using a near-field camera) is highly desirable; otherwise, intensities cannot be used reliably.

Given the fully unsupervised comparisons of diffractograms used in this work, there are numerous opportunities for novel experiment design. For example, it may be entirely unnecessary to perform both near- and far-field modalities by instead using a ``medium-field'' modality sitting between the two extremes. This particular modality is enabled by virtue of the eliminated indexing step and the unsupervised fitting approach. Further, as any number of scans can be incorporated simultaneously, it is possible to use overlapping raster scans during the reconstruction, which may further improve intragranular fidelity as each scan only includes information from the illuminated volume. This is the fundamental idea behind scanning HEDM/3DXRD \cite{hayashiScanningThreeDimensionalXRay2017}, though the present method may allow for a substantially larger beam. Rastering methods like this require precise geometric knowledge and alignment of scans, which demands additional care during experimental calibration. By combining a medium-field modality with overlapping scans, we hypothesize that the methods here can enable high-accuracy grain structure and intragranular field reconstruction within a large polycrystalline volume \emph{without} the need for point or line focusing and with a single detector.

\section*{Acknowledgements}

We are grateful to Jun-Sang Park and Peter Kenesei from the Advanced Photon Source for their contributions to the experimental data collection. We also acknowledge many fruitful discussions with Peter Voorhees and Darren Pagan, as well as with Matthew Miller, Kate Shanks, Kelly Nygren, Sven Gustafson, and Diwakar Naragani during a two-week visit to the Cornell High-Energy Synchrotron Source. We gratefully acknowledge the support of the US Office of Naval Research (No.\ N00014-21-1-2784), and the support of the US National Science Foundation through Graduate Research Fellowships to CKC and SFG (No.\ 2139433). KB also acknowledges the support of the US Office of Naval Research (MURI No.\ N00014-23-1-2654).

\section*{LLM usage statement}

Large language model assistance was used for some portions of code development. The manuscript was drafted manually without LLM use. We stand behind all statements, methods, and results as our own.

\section*{Code availability}

A GitHub repository for the source code {\texttt{PARA-X}} developed in this work will be made available publicly upon full publication.

\section*{Data availability}

The experimental data used in this work is a subset of that collected by \citeauthor{gorskeInsituVisualizationGrowing2026a}~\cite{gorskeInsituVisualizationGrowing2026a}. The specific post-processed data used in this work may be made available by request.

\appendix
\crefalias{section}{appendix}
\crefalias{subsection}{appendix}
\crefalias{subsubsection}{appendix}

\section{Forward diffraction model}\label{sec:forward_diffraction_model}

For the purposes of a self-contained representation of the forward diffraction model, we now cover the equations governing the detector position and intensity of an individual diffraction event. These expressions are mostly standard within the X-ray diffraction literature \cite{nygrenAlgorithmResolvingIntragranular2020,henningssonXrd_simulator3DXray2023,wongFrameworkGeneratingSynthetic2013}.

\subsection{Geometry}

Consider the diffraction geometry and coordinate system shown in \cref{fig:coordinate_system} with $\{e_i\}$ as the global Cartesian basis for the laboratory coordinate system. We assume that the sample is rotated exactly around $e_2$, and that the incident wave vectors are perfectly collimated (zero beam divergence) and align in the $-e_3$ direction. There are four coordinate systems to consider, defined following the naming conventions from \citeauthor{poulsenThreeDimensionalXRayDiffraction2004}~\cite{poulsenThreeDimensionalXRayDiffraction2004}:
\begin{itemize}
    \item Laboratory coordinate system $\{e_i\}$: The coordinate system in which strains and orientations are defined. The system is defined with $e_2$ as the rotational axis of the sample, $-e_3$ as the direction of the X-ray beam, and $e_1 = e_2 \times e_3$. The origin is at the intersection of the beam center and rotational axis. All other coordinate systems are defined in terms of the laboratory system.
    \item Sample coordinate system $\{e_i^\mathrm{S}\}$: transform of the laboratory system rotated about $e_2$ by $\omega$; equivalent when $\omega = 0$.
    \item Crystal coordinate system $\{e_i^\mathrm{C}\}$: coordinate system of each diffracting point based on its crystallographic orientation.
    \item Detector coordinate system $\{e_i^\mathrm{D}\}$: coordinate system with $e^\mathrm{D}_1$ and $e^\mathrm{D}_2$ aligned with the 2D detector axes and with its origin in the center of the detector.
\end{itemize}
The Bragg and azimuthal angles, $\theta$ and $\eta$, are defined in this convention as:
\begin{subequations}
    \begin{equation}
            \theta = \arcsin\left(\frac{\lambda \|g\|_2}{4\pi}\right) = \frac{1}{2}\arctan\left(\frac{\sqrt{(k_\mathrm{out})^2_1 + (k_\mathrm{out})^2_2}}{|(k_\mathrm{out})_3|}\right),
    \end{equation}
    \begin{equation}
        \eta = \arctantwo((k_\mathrm{out})_1, (k_\mathrm{out})_2),
    \end{equation}
\end{subequations}
under the standard $\arctantwo(y,x)$ ordering.

\subsection{Pointwise kinematic diffraction}\label{sec:point_wise_kinematic_diffraction}

We now consider a single point $x$ in space and the orientation and strain at that point, $q(x)$ and $\varepsilon(x)$, and explicitly define the expressions necessary to compute the detector--rotation position in which a diffraction event occurs for this point. Much of the notation and expressions follow from \citeauthor{nygrenAlgorithmResolvingIntragranular2020}~\cite{nygrenAlgorithmResolvingIntragranular2020} (see Appendix A), though with a few changes and simplifications necessary for improved symbolic gradient evaluation (particularly \cref{sec:rotation_angle}). 

\subsubsection{Orientation and deformation}

We begin by explicitly defining how the deformed reciprocal lattice vectors (RLVs) in the laboratory system are computed. We take the $\mathrm{C}$, $\mathrm{S}$, and $\mathrm{D}$ superscripts to denote that a tensor is in the crystal, sample, and detector reference frame, respectively; no superscript denotes it is in the laboratory coordinate system. Additionally, the ``0'' subscript denotes the vector is undeformed. First, consider the primitive RLVs $(b^\mathrm{C}_0)_{i}$ in the crystal reference frame. An arbitrary {RLV} $(g^\mathrm{C}_0)_\hkl$ of the $\hkl$ lattice plane in the crystal frame is then
\begin{equation}
    (g^\mathrm{C}_0)_\hkl = h (b^\mathrm{C}_0)_{1} + k (b^\mathrm{C}_0)_{2} + \ell (b^\mathrm{C}_0)_{3}.
\end{equation}
We may rotate this RLV from the crystal to the sample frame as
\begin{equation}
    (g^\mathrm{S}_0)_\hkl = Q (g^\mathrm{C}_0)_\hkl,
\end{equation}
where $Q = Q(q)$ is the (transposed) passive rotation matrix mapping vectors in the crystal frame to the sample frame \cite{rowenhorstConsistentRepresentationsConversions2015}:
\begin{equation}\label{eq:orientation_matrix}
    Q(q) = \begin{bmatrix}
        q_0^2 + q_1^2 - q_2^2 - q_3^2 & 2(q_1q_2 + q_0q_3) & 2(q_1q_3 - q_0q_2)  \\
        2(q_1 q_2 - q_0q_3) & q_0^2 - q_1^2 + q_2^2 - q_3^2 & 2(q_2q_3 + q_0q_1)   \\
        2(q_0q_2 + q_1 q_3) & 2(q_2q_3 - q_0q_1) & q_0^2 - q_1^2 - q_2^2 + q_3^2
    \end{bmatrix}.
\end{equation}
Upon a homogeneous deformation $F$ in the sample frame, the deformed RLV $g^\mathrm{S}_\hkl$ in the sample frame is then given by
\begin{equation}
    g^\mathrm{S}_\hkl = F^{-\top} Q (g^\mathrm{C}_0)_\hkl.
\end{equation}
Decomposing $F$ with left polar decomposition, then $F^{-\top} = V^{-1} R$ with $V^{-1} \approx I - \varepsilon$ where $\varepsilon = \operatorname{sym} \nabla u$. Absorbing the rotation into the crystallographic orientation yields
\begin{equation}
    g^\mathrm{S}_\hkl \approx (I - \varepsilon) Q (g^\mathrm{C}_0)_\hkl.
\end{equation}
Finally, upon rotation of the sample by $\omega$ (transformation from the sample to the laboratory frame), the deformed {RLV} in the laboratory frame is
\begin{equation}\label{eq:rlv_lab_frame}
    g_\hkl = \Omega(\omega) (I - \varepsilon) Q (g^\mathrm{C}_0)_\hkl,
\end{equation}
where $\Omega(\omega)$ is the rotation matrix which transforms from the sample to lab frame through a rotation $\omega$ around $e_2$:
\begin{equation}\label{eq:sample_rotation_matrix}
    \Omega(\omega) = \begin{bmatrix}
        \cos \omega  & 0 & \sin \omega \\
        0            & 1 & 0           \\
        -\sin \omega & 0 & \cos \omega \\
    \end{bmatrix}.
\end{equation}

\subsubsection{Rotation angle}\label{sec:rotation_angle}

We now solve for the rotation angle(s) $\omega$ under which a given lattice plane satisfies the Bragg condition. We drop the $\hkl$ subscript here and work in the laboratory frame. In this case, the diffraction can be computed from the Laue equations:
\begin{equation}\label{eq:laue_equation}
    \Delta k = k_\mathrm{out} - k_\mathrm{in} = g(\omega),
\end{equation}
where $k_\mathrm{out}$ and $k_\mathrm{in}$ are the diffracted and incoming wave vectors, and $g$ is an {RLV} from \cref{eq:rlv_lab_frame}. Assuming elastic scattering, then $\|k_\mathrm{in}\| = \|k_\mathrm{out}\|$, so
\begin{equation}\label{eq:elastic_scattering_laue}
    \|k_\mathrm{out}\| = \|g(\omega) + k_\mathrm{in}\|.
\end{equation}
Following the conventions in \cref{fig:coordinate_system}, the incident wave vector is 
\begin{equation}\label{eq:incident_wavevector}
    k_\mathrm{in} = -(2\pi / \lambda) e_3,
\end{equation}
so $\|k_\mathrm{in}\| = \|k_\mathrm{out}\| = 2 \pi / \lambda$. Expanding \cref{eq:elastic_scattering_laue} and simplifying, the following equation must hold for any elastic scattering event:
\begin{equation}\label{eq:omega_equation}
    \left[g^\mathrm{S}_1 \sin \omega - g^\mathrm{S}_3 \cos \omega\right] = -\frac{\lambda \|g^\mathrm{S}\|_2^2}{4 \pi}.
\end{equation}
The solution(s) $\omega$ to \cref{eq:omega_equation}, if they exist, are
\begin{equation}\label{eq:critical_rotation_angle}
    \omega_\pm = -\arctantwo(g^\mathrm{S}_{1}, g^\mathrm{S}_{3}) \pm \arccos\left(\frac{c}{G}\right)
\end{equation}
where $c = \lambda \|g^\mathrm{S}\|_2^2 / 4\pi$ and $G = \sqrt{(g^\mathrm{S}_{1})^2 + (g^\mathrm{S}_{3})^2}$. Note that we may use the sum and difference identities for further simplification:
\begin{subequations}\label{eq:sin_cos_omega}
\begin{equation}
    \sin \omega_\pm = \frac{-g^\mathrm{S}_{1} c \pm g^\mathrm{S}_{3} \sqrt{G^2 - c^2}}{G^2},
\end{equation}
\begin{equation}
    \cos \omega_\pm = \frac{g^\mathrm{S}_{3} c \pm g^\mathrm{S}_{1} \sqrt{G^2 - c^2}}{G^2}.
\end{equation}
\end{subequations}
We seek an angle in the range $[-\pi, \pi]$, which may be computed from
\begin{equation}
    \omega_\pm = \arctantwo(\sin \omega_\pm, \cos \omega_\pm).
\end{equation}
Given a solved $\omega_\pm$, the outgoing wave vector is computed from the Laue equations and may be expanded and simplified using the expressions in \cref{eq:sin_cos_omega} to obtain the simple expression:
\begin{equation}\label{eq:kout_final}
    k_\mathrm{out,\pm} = g(\omega_\pm) - (2\pi / \lambda) e_3 = \begin{bmatrix} \pm \sqrt{G^2 - c^2} & g^\mathrm{S}_{2} & c - 2\pi / \lambda \end{bmatrix}^\top.
\end{equation}
The laboratory-frame position the diffraction event emanates from is then $x_\pm = \Omega(\omega_\pm) x^\mathrm{S}$. Observe that \cref{eq:kout_final} clearly highlights the possible solution cases: two solutions when $c < G$, one solution when $c = G$, and no solutions when $c > G$. We may also note from \cref{eq:critical_rotation_angle} that $\omega$ becomes increasingly sensitive as $c / G \to 1$. One may extend these equations to an arbitrary rigid body motion \cite{henningssonXrd_simulator3DXray2023}, if necessary.

\subsubsection{Detector intercept}\label{sec:detector_intercept}

The detector can be tilted and offset from the laboratory origin, where the detector origin is $\zeta_0$ and the detector basis is $\{e^\mathrm{D}_1, e^\mathrm{D}_2, e^\mathrm{D}_3\}$. We take the detector origin (with reference to the laboratory origin) to be
\begin{equation}
    \zeta_0 = \begin{bmatrix} \xi_1^\mathrm{BC} & \xi_2^\mathrm{BC} & -L_\mathrm{sd}\end{bmatrix}^\top,
\end{equation}
where $\xi^\mathrm{BC}$ is the beam center (or beam offset), and $L_\mathrm{sd}$ is the distance between the sample and detector. This point corresponds to the direct center of the detector. The tilted basis is given by $e^\mathrm{D}_i = R e_i$ where $R = Q^\top(q_\mathrm{det})$ given the quaternion $q_\mathrm{det}$ describing the detector plane frame (noting $Q$ given by \cref{eq:orientation_matrix}). Taking the diffracting position $x$ and outgoing wave vector $k_\mathrm{out}$, computed in the laboratory frame in \cref{sec:rotation_angle}, the spatial position $\zeta$ at which the diffracting beam intersects the detector can be computed via the parametric representation
\begin{equation}
    \zeta = x + t k_\mathrm{out},
\end{equation}
where $t$ is the ray parameter from $x$ to $\zeta$. As the origin of the detector $\zeta_0$ lies on the detector plane and $e^\mathrm{D}_3$ is the normal to the detector, then the ray parameter is
\begin{equation}
    t = \frac{(\zeta_0 - x) \cdot e^\mathrm{D}_3}{k_\mathrm{out} \cdot e^\mathrm{D}_3}.
\end{equation}
Combining, the position where the ray intercepts the detector plane is
\begin{equation}
    \zeta = x + \frac{(\zeta_0 - x) \cdot e^\mathrm{D}_3}{k_\mathrm{out} \cdot e^\mathrm{D}_3} k_\mathrm{out}.
\end{equation}
The normalized 2D detector coordinate $\xi$ can then be computed by projecting onto the $e^\mathrm{D}_1$ and $e^\mathrm{D}_2$ detector basis as
\begin{equation}\label{eq:detector_intercept}
    \xi = \begin{bmatrix}e^\mathrm{D}_1 & e^\mathrm{D}_2\end{bmatrix}^\top (\zeta - \zeta_0) / \max(w_\mathrm{d}, h_\mathrm{d}),
\end{equation}
where $w_\mathrm{d}$ and $h_\mathrm{d}$ are the width and height of the detector, respectively. Letting $\hat w_\mathrm{d} = w_\mathrm{d} /\max(w_\mathrm{d}, h_\mathrm{d})$ and $\hat h_\mathrm{d} = h_\mathrm{d} /\max(w_\mathrm{d}, h_\mathrm{d})$, then if $\xi_1 \in [-\hat w_\mathrm{d} / 2, \hat w_\mathrm{d} / 2]$ and $\xi_2 \in [-\hat h_\mathrm{d} / 2, \hat h_\mathrm{d} / 2]$ the point appears on the detector. This form of normalization allows for non-square detectors. Following the notation described in \cref{eq:detector_rotation_space}, we define $y \coloneqq (\xi, \omega)$.

\subsubsection{Gradients}

As the expressions above are written in closed form, they may be analytically differentiated using symbolic tools\footnote{The spot position is only piece-wise smooth as spot solutions may cease to exist or spots may leave or enter the measured detector--rotation domain.}; we use SymPy~\cite{meurerSymPySymbolicComputing2017} to compute such gradients. To reduce the expression length of analytic derivatives, the simplest form of each expression being differentiated is used. We compute ${\partial y}/{\partial (\cdot)}$, where again $y\coloneqq (\xi,\omega)$ and $(\cdot)$ are $q$, $\varepsilon$, and $x^\mathrm{S}$. The spot position is also a function of several parameters that are held constant during the experiment, including, but not limited to, the detector tilts, beam center, and distance; X-ray energy; and lattice parameters. Gradients with respect to these parameters may also be computed and can be used for calibration purposes, although these are generally not the values we wish to measure during an HEDM experiment. We note here that recent work evaluates these derivatives using autograd \cite{sharmaEndtoEndDifferentiableForward2026}; this is another possibility.

\subsection{Diffraction intensity}\label{sec:intensity}

We now outline the main components involved within the intensity of a given Bragg reflection. The intensity of a given $\hkl$ reflection may be computed as 
\begin{equation}\label{eq:diffraction_intensity}
    I_\hkl(q,\varepsilon,x) \propto V I^0_\hkl A_\hkl|F_\hkl|^2 L_\hkl P_\hkl,
\end{equation}
where the proportionality is dependent on factors that are assumed to be held constant during an experiment (e.g., incident photon flux and energy, rotation speed, exposure time, etc.), $V$ is the diffracting volume, $I^0_\hkl$ and $A_\hkl$ are, respectively, the incident flux and ray attenuation at the location of the diffraction event, and $F_\hkl$, $L_\hkl$, and $P_\hkl$ are the structure, Lorentz, and polarization factors, respectively. Although the intensity depends on the strain, orientation, and position, we make the simplifying assumption that $\partial I_\hkl / \partial q = \partial I_\hkl / \partial \varepsilon \approx 0$ in all subsequent expressions. This is only a poor approximation when the azimuthal angle is near a pole and the Lorentz factor becomes singular, but these reflections are masked anyway. Note here that many expressions are given up to a constant factor; within all intensity-modulating expressions, constants are assumed to be contained within the scaling parameter $\beta$ in \cref{eq:discrete_probability_measures}. 

\subsubsection{Structure factor}

The form of the structure factor $F_\hkl$ is
\begin{equation}
    F_\hkl = \sum_{j=1}^N C_j f_j T_j \exp \left[i g_\hkl \cdot r_j \right],
\end{equation}
where $C_j$, $f_j$, and $T_j$ are the site-occupancy, atomic scattering, and Debye--Waller factors of atom $j$, respectively, with unit cell position $r_j$. The atomic scattering factor of a given atom is the sum of the Thompson $f_0$, resonant $f' + if''$, and nuclear Thomson $f_\mathrm{NT}$ scattering factors:
\begin{equation}
    f(g_\hkl, \lambda) = f_0 + f' + if'' + f_\mathrm{NT}.
\end{equation}
We compute values for $f_0$ from the Gaussian mixture coefficients given by \citeauthor{waasmaierNewAnalyticalScatteringfactor1995}~\cite{waasmaierNewAnalyticalScatteringfactor1995} and $f'$ and $f''$ from the tabulated data by \citeauthor{kisselRTABRayleighScattering2000}~\cite{kisselRTABRayleighScattering2000}. The Debye--Waller factor $T$ accounts for coherent scattering attenuation due to thermal vibrations in the atom and is given, assuming isotropic atomic displacement, by
\begin{equation}
    T_j = \exp\left(\frac{-2 \pi^2 U_j}{d_\hkl^2}\right),
\end{equation}
where $U_j = \langle u^2\rangle$ is the isotropic mean square displacement of atom $j$, and $d_\hkl$ is the lattice plane spacing. The atomic scattering factors are computed for the unstrained lattice and assumed to be unchanged through deformation (i.e., the structure factor is computed once).

Exact knowledge of the structure factor is less significant for far-field comparisons as all reflections from the same family have the same structure factor (neglecting deformation); however, near-field comparisons use all families simultaneously. Therefore, for grain boundary updates, refinement of the crystal structure via powder X-ray diffraction is necessary, particularly for complex crystal systems. For the example AlON sample used in this work, a separate Rietveld structure refinement was performed using powder X-ray diffraction data collected from the raw material stock.

\subsubsection{Geometric factors}

\paragraph{Lorentz factor}

The Lorentz factor for a rotating crystal (expressed in the coordinate system in \cref{fig:coordinate_system}) is \cite{milchIndexingSinglecrystalXray1974}
\begin{equation}\label{eq:general_lorentz_factor}
    L^{-1} \propto |\hat k_\mathrm{out} \cdot (\dot \omega {e}_2 \times g_\hkl)|,
\end{equation}
where $\dot \omega e_2$ is the angular velocity, and the hat denotes the vector is unit-normalized. Inserting the Laue equation from \cref{eq:laue_equation}, the Lorentz factor for this system may be expressed (up to a constant) as
\begin{equation}
    L^{-1} = |\hat k_\mathrm{out} \cdot ({e}_2 \times (\hat k_\mathrm{out} - \hat k_\mathrm{in}))| = |{e}_2 \cdot (\hat k_\mathrm{in} \times \hat k_\mathrm{out})|.
\end{equation}
Finally, the incident wave vector in \cref{eq:incident_wavevector} is aligned in the $-\hat{e}_3$-direction, so the Lorentz factor may be expressed in its final form as
\begin{equation}\label{eq:final_lorentz_factor}
    L \approx \frac{1}{|(\hat k_\mathrm{out})_1| + \delta}.
\end{equation}
A small positive constant $\delta$ is added to \cref{eq:final_lorentz_factor} to prevent reflections from contaminating the overall diffractogram comparison when the Lorentz factor becomes singular at the azimuthal poles; this allows $L$ to be differentiable if intensity gradients are desired. In this work, $\delta = 10^{-2}$, so the Lorentz factor is restricted from increasing beyond $L=100$. The form in \cref{eq:final_lorentz_factor} (without the numerical constant) is equivalent to that given in terms of $2\theta$ and $\eta$ by \citeauthor{lauridsenTrackingMethodStructural2001}~\cite{lauridsenTrackingMethodStructural2001}.

\paragraph{Polarization factor}

Given arbitrary polarization states of the incident and outgoing rays, ${\hat p}_\mathrm{in}$ and $\hat{p}_\mathrm{out}$, respectively, the scattering cross-section of a single scattering event is \cite{als-nielsenElementsModernXray2017}
\begin{equation}
    \frac{\mathrm{d}\sigma}{\mathrm{d}\Omega} \propto |{\hat p}_\mathrm{in} \cdot {\hat p}_\mathrm{out}|^2.
\end{equation}
Assuming that the beam is partially linearly polarized with fraction $p_\mathrm{h}$ in the horizontal $e_1$ direction, the polarization factor for any diffraction event may be computed as 
\begin{equation}
    P = p_\mathrm{h}\left[1 - (\hat k_{\mathrm{out}})_1^2\right] + (1 - p_\mathrm{h})\left[1 - (\hat k_{\mathrm{out}})_2^2\right].
\end{equation}
In this work, we assume for simplicity that the incident beam is entirely horizontally polarized: $p_\mathrm{h} = 1$ as synchrotron X-rays are strongly horizontally polarized. 

\subsubsection{Instrument}

We now outline other factors which influence the intensity. Note that we explicitly do not discuss point spread here; its implementation, including how gradients are propagated, is described in \cref{sec:point_spread}.

\paragraph{Attenuation}

The attenuation of the incident and diffracted beam may be approximately accounted for by ray tracing the path the beam takes through the sample and assuming a uniform attenuation coefficient and ideal sample geometry. The vertical boundaries of the sample are first approximated with a set of planes (four for the case of a hexahedral sample). For each diffraction event, the path length through the sample, $\ell$, is computed by summing the minimum intersection distance between the incoming and outgoing wave vectors and each plane. This computation is performed with the sample rotated by the rotation angle $\omega$ of the diffraction event. The attenuation of the beam may then be computed by the Beer--Lambert law:
\begin{equation}
    A = \exp(-\mu  \ell),
\end{equation}
where $\mu$ is the linear attenuation coefficient of the material. The attenuation may be neglected in many cases depending on the energy, sample size, or elemental composition. Attenuation due to air is neglected.

\paragraph{Incident beam}

Spatiotemporal variation in intensity and energy of the beam may be accounted for through some standard and non-standard measurements. Temporal variations may be accounted for using ion chamber measurements normally taken at synchrotron facilities. For intensity, at each imaging frame, the total flux of the incident beam is proportional to the counts in the final ion chamber after all beam conditioning. The intensity fluctuation is accounted for by inversely scaling each frame of reference experimental data by the associated ion chamber measurement; this is easier than scaling each simulated event. Temporal energy fluctuations may be approximately accounted for through foil attenuation measurements that are normally taken for calibration of the monochromator energy. Fully incorporating time-varying energy would require a fixed-point iteration as the rotation increment is dependent on the energy. The spatial intensity variation of the beam may be measured directly with the near-field camera using a long enough exposure time to get a time-averaged spatial distribution where the spatial intensity distribution can be assumed constant. For each diffraction event, the intensity is computed by sampling the beam profile at the rotated point $x$.

In this initial work, we only consider the approximate spatial variation in intensity of the beam. We do not consider temporal intensity changes, or any spatiotemporal energy changes. In future work, these factors should be considered by the methods outlined above.

\paragraph{Other}

Beyond the point-spread implementation discussed in \cref{sec:point_spread}, there are numerous other factors that influence the intensity of a diffraction event. Such factors include, but are not limited to, detector solid-angle or other non-idealities, beam divergence and/or energy spread, and dynamical diffraction. It remains to be seen which factors are necessary to incorporate into the forward model approach developed in this work.

\subsection{Forward diffraction function}

Given a kinematically computed spot position $y_\hkl \in \mathcal Y$ and associated intensity $I_\hkl$, we now define the forward diffraction function used in this work. At each point, all lattice planes can possibly diffract; however, we search over $\hkl \in \mathcal H$, where $\mathcal H$ is the space of admissible lattice planes, which we define as:
\begin{equation}\label{eq:admissible_hkls}
    \mathcal H \coloneqq \{\hkl \in \mathbb Z^3 \mid |F_\hkl|^2 > F_\mathrm{min}, \ \theta_\hkl < \theta_\mathrm{max}\},
\end{equation}
where $F_\mathrm{min}$ is some minimum structure factor magnitude (2\% of the maximum value), and $\theta_\mathrm{max}$ is the maximum Bragg angle to consider (chosen to be the maximum possible detectable angle). Given the admissible lattice planes in $\mathcal H$, and the two possible solutions $\omega_\pm$ for each plane, the forward diffraction function for a single point is then defined as
\begin{equation}\label{eq:diffraction_forward_function}
    f(y, x, q(x), \nabla u(x)) \coloneqq \sum_{\hkl \in \mathcal H} \sum_{\omega_\pm} I_{hk\ell}\Bigl(x, q(x), \nabla u(x)\Bigr)\delta\Bigl(y - y_\hkl\bigl(x, q(x), \nabla u(x)\bigr)\Bigr).
\end{equation}
Dropping the $x$ dependencies for succinctness, the partial derivatives of \cref{eq:diffraction_forward_function} are
\begin{equation}\label{eq:partial_derivatives_of_forward}
    \frac{\partial f}{\partial (\cdot)}(y) = \sum_{\hkl \in \mathcal H} \sum_{\omega_\pm} \left(\delta(y - y_\hkl)\frac{\partial I_\hkl}{\partial (\cdot)} - I_\hkl \nabla_y\delta(y - y_\hkl) \cdot \frac{\partial y_\hkl}{\partial (\cdot)} \right),
\end{equation}
where $(\cdot)$ is $x$, $q$, or $\varepsilon$, and the Dirac delta function~$\delta$ and its gradient~$\nabla_y \delta$ are defined in the usual weak sense, and only have meaning when integrated over $\mathcal Y$. Again note that we assume ${\partial I_\hkl} / {\partial (\cdot)} \approx 0$ in this work as \cref{eq:diffraction_intensity} is difficult to differentiate and the gradients are small for most points.

\subsection{Polycrystal diffraction}

For the purposes of generating full virtual diffractograms from an entire illuminated section of the polycrystal, the forward diffraction function is numerically integrated over each element in the mesh\footnote{It is also possible to numerically integrate over the incident energy distribution, imparted due to non-ideal monochromation, at the expense of computational cost.}:
\begin{equation}
    I(y) \approx \sum_E \sum_{x \in \bar E} f(y, x, q(x),\nabla u(x)),
\end{equation}
where here $E$ is a given element within the mesh, and $\bar E$ is a collection of points within the element that are contained within the beam \cite{wongFrameworkGeneratingSynthetic2013} (note that we take $V$ in \cref{eq:diffraction_intensity} to be the illuminated fraction of the element for a given reflection). We usually do not use the finite element Gauss points for the numeric integration as near-field data generally require finer spatial resolution than the element spacing. This also improves integration for partially illuminated elements. As the strain and orientation in each element are assumed constant, we compute the intensity $I_\hkl$, the outgoing wave vector $k_\mathrm{out}$, and the rotation $\omega$ for each $\hkl \in \mathcal H$ over the element, all of which are constant with respect to the position. We then subdivide each element using $n$ points along each edge and compute the global coordinate $x^\mathrm{S}$ associated with each point in the subdivision. The spot position $y \in \mathcal Y$ and intensity from each subdivision point may then be computed using the above quantities by following \cref{sec:detector_intercept}. In the case of synthetic data generation, the diffractogram is discretized over the detector--rotation space (discrete pixels and finite rotation increments), where the intensity at a single pixel $P \subset [-0.5, 0.5]^2$ over rotation increment $W \subset [-\pi, \pi]$ is simply
\begin{equation}
I(P, W) = \int_{P \times W} \int_\Omega f (y, x, q(x), h(x)) \dif x \dif y.
\end{equation}

\section{Further computational details}\label{sec:further_computational_details}

There are several additional modeling choices we make that can improve the efficacy of the methods presented in this work. We find these implementation details useful, though whether they are strictly necessary depends on the experimental conditions. 

\subsection{Transport space}

The underlying space over which the optimal transport distance is computed has significant consequences for how well the method performs, and as there is nothing fundamental about the space $\mathcal Y$, we are free to transform it as desired\footnote{The optimal transport distance weights points by their mass, so spots with higher intensity are effectively given more weight during gradient computation. Additional transformation of the intensity (e.g., log scaling) may be desired, but we do not do so here.}. In numerical experiments, we found it advantageous, though not necessary, to represent spots in the form $\hat y = (\hat r, \hat \eta, \hat \omega) \in \hat {\mathcal Y}$ when evaluating the optimal transport distance, where
\begin{equation}
    \hat {\mathcal Y} \coloneqq \{(\hat r, \hat \eta, \hat \omega) \mid \hat r \in [0, 1], \ \hat \eta \in [-0.5, 0.5], \ \hat \omega \in [-0.5, 0.5]\}.
\end{equation}
The change of variables $y \in \mathcal Y \mapsto \hat y \in \hat {\mathcal Y}$ may then be defined as
\begin{subequations}
    \begin{equation}
        \hat r \coloneqq \sqrt{2(\xi_1 - \xi_1^\mathrm{BC})^2 + 2(\xi_2 - \xi_2^\mathrm{BC})^2},
    \end{equation}
    \begin{equation}
        \hat \eta \coloneqq \arctantwo(\xi_1 - \xi_1^\mathrm{BC}, \xi_2 - \xi_2^\mathrm{BC}) / 2\pi,
    \end{equation}
    \begin{equation}\label{eq:rotation_change_of_variables}
        \hat \omega \coloneqq \omega / 2\pi,
    \end{equation}
\end{subequations}
where the choice of scaling in $\hat r$ allows values $\hat r \in [0, 1]$. The corresponding gradient transformation is
\begin{equation}
    \frac{\partial d}{\partial y_{i}} = \frac{\partial d}{\partial \hat y_{j}}\frac{\partial \hat y_{j}}{\partial y_{i}},
\end{equation}
where
\begin{subequations}
\begin{equation}
    \frac{\partial \hat r}{\partial \xi_1} = 2 \frac{\xi_1 - \xi_1^\mathrm{BC}}{\hat{r}}, \quad \frac{\partial \hat r}{\partial \xi_2} = 2\frac{\xi_2 - \xi_2^\mathrm{BC}}{\hat{r}},
\end{equation}
\begin{equation}
    \frac{\partial \hat \eta}{\partial \xi_1} = \frac{\xi_2 - \xi_2^\mathrm{BC}}{2\pi ((\xi_1 - \xi_1^\mathrm{BC})^2 + (\xi_2 - \xi_2^\mathrm{BC})^2)}, \quad \frac{\partial \hat \eta}{\partial \xi_2} = -\frac{\xi_1 - \xi_1^\mathrm{BC}}{2\pi ((\xi_1 - \xi_1^\mathrm{BC})^2 + (\xi_2 - \xi_2^\mathrm{BC})^2)},
\end{equation}
\begin{equation}\label{eq:rotation_gradient_change_of_variables}
    \frac{\partial \hat \omega}{\partial \omega} = 1 / 2\pi.
\end{equation}
\end{subequations}
This change of variables accentuates radial discrepancies in diffracted positions, which are primarily caused by deformations in the diffracting lattice plane. For near-field data, we perform a change of variables only for the rotation following \cref{eq:rotation_change_of_variables,eq:rotation_gradient_change_of_variables}.

For far-field data, we perform one additional reduction: split the objective function into distance calculations over each Debye--Scherrer ring. It is clear from the diffraction topology that all spots lying on a given Debye--Scherrer ring must come from a single family (or families) of known lattice planes, so it is unnecessary to consider interactions between points on distinct rings. In this sense, we split the objective function into a sum of smaller problems over each ring as 
\begin{equation}
    d = \frac{\sum_{i} m_i d_i}{\sum_{i} m_i},
\end{equation}
where $m_i$ is the multiplicity of the $i$th ring. The multiplicity is added to the weighting as we also unit-normalize the intensities over each ring. The multiplicity scaling re-adds the natural weighting by the total number of diffraction points appearing from a given point in space. This splitting procedure also lessens the need for precise knowledge of the structure factor of a given plane family as it may be normalized out during the forward diffraction computation. Splitting the diffractogram objective evaluation is possible only when the detector is far enough that the Debye--Scherrer rings are distinct.

\subsection{Masking}

The equations that govern diffraction position and intensity have a singularity when the azimuthal angle $\eta$ (see \cref{fig:coordinate_system}) of the diffraction event approaches zero. First, the solution to the rotation at which the diffraction event occurs (see \cref{eq:critical_rotation_angle}) becomes extremely sensitive to the orientation of the crystal, so reflections can become smeared across many rotation frames, especially so if there is large mosaic spread or strain gradients within the crystal. Second, the Lorentz factor approaches infinity as the azimuthal angle approaches a pole: $\lim_{\eta \to 0} L = \lim_{\eta \to \pm 180^\circ} L = \infty$. In practice, these two factors mean that such reflections, henceforth called ``near-singular reflections,'' can decrease the effectiveness of the overall approach presented here due to non-idealities in the experimental conditions and forward diffraction model.

To prevent these near-singular reflections from degrading performance, the reflections are masked in one of two ways. In the case of far-field data, the reflections all lie near the distinct Debye--Scherrer rings, so masking these spots is straightforward: the azimuthal angle of each pixel is computed (assuming a reflection originating from $x = 0$) and pixels with
\begin{equation}\label{eq:eta_mask_range}
    \min(|\eta|, 180^\circ - |\eta|) \leq \eta_\mathrm{mask}
\end{equation}
are masked such that any simulated or experimental points that hit such a pixel are discarded. In the near-field case, near-singular reflections are harder to neglect as the azimuthal angle of experimental reflections is unknown, so we cannot discard data from the diffractograms a priori. As a result, for near-field data, we do not mask by azimuthal angle in the forward diffraction computation, and instead neglect gradient updates from reflections whose azimuthal angle lies in the same masked range in \cref{eq:eta_mask_range}. In this way, the near-singular reflections are included during the calculation of the optimal transport distance, but their gradients are neglected. Masking is applied, however, to discard detector points near the direct beam or blocked by the near-field beam stop.

\subsection{Point spread}\label{sec:point_spread}

To incorporate point spread in the simulated diffractograms, a 2D discrete Gaussian point spread stencil is placed at the center of the main computed spot. The point spread stencil consists of finite weights computed by integrating the Gaussian over the stencil discretization and normalizing. Perturbing the main point perturbs the stencil points equally; therefore, each point can be assigned the same gradients as the main spot. In this sense, the gradient is only computed once and copied to each sub-point within the point spread stencil. This method may be applied to other point spread functions, so long as a discretized stencil is used.

From synthetic experiments, accurate point spread incorporation seems to be only necessary from the standpoint of intensity thresholding. In the case that diffraction events are treated as Dirac masses, some binned diffraction spots, particularly in the near-field case, may be sparsely populated due to the finite sampling of points within an element. In this sense, point spread enables more accurate thresholding of spots near the detector noise floor.

\subsection{Normalization}\label{sec:normalization}

The intensity scaling factor $\beta$ in \cref{eq:discrete_probability_measures} is initially set such that $\nu_\mathrm{sim}$ is a probability measure:
\begin{equation}\label{eq:initial_alpha}
    \beta = \left[\sum_{i=1}^n I_i\right]^{-1}.
\end{equation}
The reference experimentally measured intensity distribution is also normalized in the same way. We then perform a grid search over 25 points by varying $\beta$ within 25\% of its initial value and choose the $\beta$ that minimizes the objective; this is the initial guess for the first iteration. This method allows for flexibility if the threshold intensity of the detector is too high or low. After each subsequent optimization iteration, we update $\beta$ by computing
\begin{equation}
    \frac{\mathrm{d}d}{\mathrm{d}\beta} = \sum_{i} I_i \frac{\mathrm{d}d}{\mathrm{d}I_i}, 
\end{equation}
where here $I_i$ is the intensity of the $i$th binned spot. We then update $\beta$ with standard gradient descent. This empirically leads to smoother objective function decay as opposed to per-iteration normalization of $I$ to a probability measure. The normalization constant is updated individually for each distinct objective function evaluation, so it is computed on a per-ring basis for each far-field layer, and on a per-layer basis for near-field data.

\subsection{Update modalities}\label{sec:update_modalities}

One may use either near- or far-field data for any of the local field or grain boundary updates, either simultaneously or individually restricted to one modality. Near-field diffractograms, by experimental definition, contain significantly more spatial information, so near-field modalities are the best choice for grain boundary advection. Far-field data are more limited, in that the grain size is encoded within the total intensity of diffraction spots, but the local grain topology is not. Due to their strain sensitivity, far-field data are better suited for traction updates following \cref{sec:pde_constrained_point_wise}, and may be used for orientation field updates should there be low intragranular mosaicity. In this work, we do not use both near- and far-field modalities simultaneously for all updates, though this is possible under sufficient alignment and objective function scaling (gradient magnitudes may vary by orders of magnitude between the two modalities). In some cases, it is also prudent to perform orientation averaging over each grain for some number of initial iterations, depending on the material.

\section{Traction constraint}\label{sec:traction_discrete_constraint}

We now outline the discrete implementation of the traction constraint in \cref{eq:self_equilibrium}. Upon finite element discretization, we consider each surface element to have a constant traction over the external face(s). Given this, the discrete form of the space of admissible tractions in \cref{eq:self_equilibrium} is $Ct_\mathrm{s} = f_0$ where $C \in \R^{7 \times 3n}$ and $t_\mathrm{s} \in \R^{3n}$ with $n$ total surface element faces with tractions applied. Each row of $C$ computes the net contribution to the force/moment due to applied traction on the face. For each external face with area $A_\mathrm{loc}$, we assemble the local $C_\mathrm{loc} \in \R^{7 \times 3}$ matrix, which computes the force and moment of each face, and the net force vector as
\begin{equation}
    C_\mathrm{loc} = 
\begin{bmatrix}
I_\mathrm{t} + I_\mathrm{b} & 0 & 0 \\
0 & I_\mathrm{t} & 0 \\
0 & I_\mathrm{b} & 0 \\
0 & 0 &  I_\mathrm{t} + I_\mathrm{b}\\
0 & -I_z & I_y \\
I_z & 0 & -I_x \\
-I_y & I_x & 0
\end{bmatrix}, \quad
f_0 = 
\begin{bmatrix}
0 \\ F_\mathrm{net} \\ -F_\mathrm{net} \\ 0 \\ 0 \\ 0 \\ 0
\end{bmatrix},
\end{equation}
where the terms are given by the local area integrals:
\begin{equation}
    I_\mathrm{t} = \int_{A^\mathrm{top}_\mathrm{loc}} \dif A, I_\mathrm{b} = \int_{A^\mathrm{bot}_\mathrm{loc}} \dif A, I_x = \int_{A_\mathrm{loc}} x_1 \dif A,  I_y = \int_{A_\mathrm{loc}} x_2 \dif A,  I_z = \int_{A_\mathrm{loc}} x_3 \dif A,
\end{equation}
which are evaluated in standard fashion. Using surface tractions here is preferred over nodal forces as the constraint is mesh-size independent (e.g., see Ref.~\cite{cockeRecoveringIntragranularStrain2025}). Given a computed $C$, the projection onto the net-force- and moment-free subspace is $P = I - QQ^\top$ where $Q \in \R^{3n \times 7}$ is the orthonormal matrix from thin QR-decomposition of $C^\top$. We then use standard projected gradient descent to update the traction distribution. The uniform traction initial guess outlined in \cref{sec:initial_guess} satisfies $Ct=f_0$ if the top and bottom surfaces are identical up to a vertical translation (the case in this work).

\printbibliography

\end{document}